\documentclass[a4paper]{emulateapj} \newcommand{\ltab}{\LongTables}

\usepackage{hyperref}

\hypersetup{
    bookmarks=true,         
    unicode=false,          
    pdftoolbar=true,        
    pdfmenubar=true,        
    pdffitwindow=true,     
    pdfstartview={FitH},    
    pdftitle={cemp stars from HES},    
    pdfauthor={vmplacco},     
    pdfsubject={Astronomy},   
    pdfcreator={dvipdf},   
    pdfproducer={dvipdf}, 
    pdfkeywords={metal-poor stars},
    pdfnewwindow=true,      
    colorlinks=true,       
    linkcolor=red,          
    citecolor=blue,        
    filecolor=magenta,      
    urlcolor=cyan,           
    breaklinks=true,
    linktocpage
}

\newcommand{\metal}{[Fe/{}H]}
\newcommand{\cfe}{[C/{}Fe]}
\newcommand{\abund}[2]{[#1/{}#2]}
\newcommand{\cemp}{{CEMP}}
\newcommand{\teff}{$T_{\rm eff}$\,}
\newcommand{\logg}{log\,$g$\,}

\newcommand{\jk}{({J$-$K})$_0$}
\newcommand{\sstar}{HE~2138$-$3336}
\newcommand{\rstar}{HE~2258$-$6358}

\shorttitle{Metal-Poor Stars Observed with the Magellan Telescope I}
\shortauthors{Placco et al.}

\begin{document}

\title{Metal-Poor Stars Observed with the Magellan Telescope. I\footnotemark[1].\\
Constraints on Progenitor Mass and Metallicity of AGB Stars \\ Undergoing s-Process
Nucleosynthesis}

 \footnotetext[1]{Based on 
observations gathered with the 6.5 meter Magellan Telescopes 
located at Las Campanas Observatory, Chile.}

\author{Vinicius M. Placco}
\affil{Departamento de Astronomia - Instituto de Astronomia, 
Geof\'isica e Ci\^encias Atmosf\'ericas, Universidade de S\~ao Paulo, \\
S\~ao Paulo, SP 05508-900, Brazil}

\author{Anna Frebel}
\affil{Massachussetts Institute of Technology and Kavli Institute 
for Astrophysics and Space Research, 77 Massachusetts Avenue, \\
Cambridge, MA, 02139, USA}

\author{Timothy C. Beers}
\affil{National Optical Astronomy Observatory, Tucson, AZ 85719, USA}

\author{Amanda I. Karakas, Catherine R. Kennedy}
\affil{Research School of Astronomy and Astrophysics, 
The Australian National University, Cotter Road, Weston, ACT, 2611, Australia}

\author{Silvia Rossi}
\affil{Departamento de Astronomia - Instituto de Astronomia, 
Geof\'isica e Ci\^encias Atmosf\'ericas, Universidade de S\~ao Paulo, \\
S\~ao Paulo, SP 05508-900, Brazil}

\author{Norbert Christlieb}
\affil{Zentrum f\"ur Astronomie der Universit\"at Heidelberg, 
Landessternwarte, K\"onigstuhl 12, 69117, Heidelberg, Germany}

\author{Richard J. Stancliffe}
\affil{Argelander-Institut f\"ur Astronomie der Universit\"at Bonn, Auf 
dem H\"ugel 71, 53121, Bonn, Germany}

\begin{abstract}

We present a comprehensive abundance analysis of two newly-discovered
carbon-enhanced metal-poor (CEMP) stars. \sstar{} is a $s$-process-rich
star with $\mbox{[Fe/H]} = -2.79$, and has the highest [Pb/Fe] abundance
ratio measured thus far, if NLTE corrections are included
($\mbox{[Pb/Fe]} = +3.84$).  \rstar{}, with $\mbox{[Fe/H]} = -2.67$,
exhibits enrichments in both $s$- and $r$-process elements. These stars
were selected from a sample of candidate metal-poor stars from the
Hamburg/ESO objective-prism survey, and followed up with
medium-resolution ($R\sim 2,000$) spectroscopy with GEMINI/GMOS. We
report here on derived abundances (or limits) for a total of 34 elements
in each star, based on high-resolution ($R\sim 30,000$) spectroscopy
obtained with Magellan-Clay/MIKE. Our results are compared to
predictions from new theoretical AGB nucleosynthesis models of 1.3\,
M$_{\odot}$ with [Fe/H] = $-$2.5 and $-$2.8, as well as to a set of AGB
models of 1.0 to 6.0\, M$_{\odot}$ at [Fe/H] = $-$2.3. The agreement
with the model predictions suggests that the neutron-capture material in
\sstar{} originated from mass transfer from a binary companion star that
previously went through the AGB phase, whereas for \rstar, an additional
process has to be taken into account to explain its abundance pattern.
We find that a narrow range of progenitor masses (1.0$\leq$\,
M(M$_{\odot}$)\,$\leq$1.3) and metallicities ($-2.8\leq$\, \metal\,
$\leq-2.5$) yield the best agreement with our observed elemental
abundance patterns. 

\end{abstract}

\keywords{Galaxy: halo---methods: spectroscopy---stars: 
abundances---stars: atmospheres---stars:
  Population II}

\section{Introduction}
\label{intro}

Chemical abundances for very metal-poor (VMP; [Fe/H $< -2.0$)
stars provide the basis for the study of the characteristic
nucleosynthetic signatures of the first few stellar generations. While
the origin of the lighter elements up to and including the iron-peak
elements are reasonably well-modeled in terms of core-collapse
supernovae (SN) nucleosynthesis \citep[e.g., ][]{woosley1995,
nomoto2006}, the production of neutron-capture elements is more complex,
and likely occurs in a range of different astrophysical sites \citep[see
e.g.,][and references therein]{sneden2008}.

The {\it{slow}} neutron-capture process \citep[$s$-process; ][]{b2fh}
has been confirmed theoretically and observationally to occur in
thermally-pulsing (TP) asymptotic giant-branch stars 
\citep[AGB; e.g.,][]{smith1987,smith1990,busso2001,abia2002}. 
AGB nucleosynthesis predictions are subject to many important
uncertainties, including the treatment of convection, which determines
the level of chemical enrichment due to the mixing of nuclear-processed
material from the core to the envelope, as well as mass loss, which determines
the AGB lifetime. The operation of the $s$-process in AGB stars also
depends on the formation of a $^{13}$C ``pocket'' for efficient
activation of the $^{13}$C($\alpha$,n)$^{16}$O neutron-producing
reaction \citep[e.g.,][]{busso1999}. The formation, shape, and the
extent in mass of the helium intershell region of such $^{13}$C pockets
is unknown, and highly uncertain \citep[see discussions
in][]{cristallo2009, bisterzo2010,lugaro2012}. As a result, accurate
elemental-abundance observations provide the best constraint on the
stellar models, enabling stringent tests of the nucleosynthesis
predictions. Massive stars also produce some $s$-process elements, with
the most recent models suggesting that their contribution is especially
important at the lowest metallicities \citep[e.g.,][]{pignatari2010,
frischknecht2012}.

While the $s$-process presents a well-established framework, the
{\it{rapid}} neutron-capture process ($r$-process) has proven more
challenging in terms of experimental determinations, due to the
difficulty in observing the properties of the isotopes involved in this
process. The $r$-process requires large neutron number densities to
occur, and this condition argues for explosive environments, such as
supernova explosion, or neutron star and black hole mergers,
accretion-induced collapse models, among others 
\citep[see][and references therein]{sneden2008}.
However, even with clear observational evidence of the operation of 
the $r$-process in metal-poor stars \citep[][among others]{sneden2003,barklem2005},
these models are not yet successful in reproducing 
the abundance distribution of $r$-process elements found in stellar atmospheres.

VMP stars with clear enrichments of carbon
\citep[\cfe $\ge +1.0$; ][]{beers2005,aoki2008} are of particular
interest in this regard. Most of these carbon-enhanced metal-poor (CEMP)
stars \citep[80\% according to][]{aoki2007} exhibit the presence of
heavy elements produced by the $s$-process \citep[\cemp-s stars;
][]{beers2005}. Qualitatively, the origin of
\cemp-s stars is consistent with the hypothesis that the carbon and
$s$-process elements are due to nucleosynthesis processes that took
place during the AGB stage of evolution. In most cases, enrichment took
place in a wide binary system where the progenitor AGB star has long ago
evolved to become a white dwarf \citep{stancliffe2008}, although there
is at least one case where the CEMP star is possibly now in the TP-AGB
phase \citep{masseron2006}.

There also exist a handful of CEMP stars known with enrichments in
$r$-process elements, as well as many that exhibit both $s$- and
$r$-process element enhancements. The possible origins of the abundance
patterns of the latter class (CEMP-r/s) are currently a source of debate
in the literature, since they cannot be explained by conventional
s-process production in AGB star models.
\citet{jonsell2006} suggest a number of possible scenarios for the 
occurance of the CEMP-r/s stars, including a r-process pre-enriched 
(from pollution by Type II supernovae) molecular cloud from which the binary 
system was formed \citep[this scenario is also suggested by][]{bisterzo2009}.
However, more statistics on these objects must be gathered in order to see
whether all the CEMP-r/s can be explained by the same formation scenario.

In this work we present an elemental-abundance analysis of two 
newly-discovered \cemp{} stars, and compare their observed patterns with
yields from stellar evolution models \citep[e.g., those presented here
and by][]{lugaro2012}. This comparison is important to understand the
operation of the $s$-process at low metallicity, and to further
constrain the onset of $s$-process nucleosynthesis in the Galaxy, as
well as different stellar and Galactic chemical-evolution scenarios
\citep[e.g.,][]{hirschi2007}.  

This paper is outlined as follows. Section \ref{secobs} describes details
of the target selection, as well as the medium- and high-resolution
spectroscopic observations. The determination of stellar parameters from
the high-resolution spectroscopy, and a comparison with the
medium-resolution values, are presented in Section \ref{params},
followed by the detailed abundance analysis described in Section
\ref{abundsec}. A discussion of the elemental-abundance patterns of
these stars, and comparisons with model predictions based on $s$-process
nucleosynthesis, are presented in Section \ref{discuss}. Our conclusions
and perspectives for future work are given in Section \ref{final}.

\section{Target Selection and Observations}
\label{secobs}

The target selection and observations were carried out in three main
steps. First, visual inspection of low-resolution (R~$\sim300$) spectra
from the Hamburg/ESO Survey \citep[HES; ][]{christlieb2003} for a
carefully-selected set of CEMP candidate stars was
carried out, in order to eliminate objects with spectral flaws or other
peculiarities. Secondly, follow-up medium-resolution (R~$\sim2,000$)
spectroscopy with the Gemini-S telescope was obtained (in queue mode,
during poor observing conditions), enabling estimates of the stellar
parameters and carbon abundances. Finally, the most promising targets,
i.e., the most metal-poor stars, were observed at high spectral
resolution (R~$\sim30,000$) with the Magellan-Clay telescope, in order
to determine the chemical abundances for many elements, and to establish
their detailed abundance patterns. Details of each step are provided
below.

Although the HES was initially designed for discovering faint
extragalactic quasars \citep{reimers1990,wisotzki2000}, the HES spectra
(resolution of 15\,{\AA}, at Ca\,{\sc{ii}}\,K, and wavelength coverage
of 3200-5300\,{\AA}) have been used for searching for different types of
objects, in particular large numbers of metal-poor stars in the Galaxy.
The discoveries from the HES stellar database include the two most
iron-poor stars found to date: HE~0107$-$5240 \citep[\metal=$-$5.2,
][]{christlieb2002}, and HE~1327$-$2326, \citep[\metal=$-$5.4,
][]{frebel2005,aoki2006}. A number of other searches
have been carried out, aiming to find, e.g., carbon-rich stars
of all metallicities \citep{christlieb2001}, field horizontal-branch
stars \citep{christlieb2005}, and bright metal-poor stars
\citep{frebel2006}.

The \cemp~ star candidates presented in this work were selected on the
basis of their strong molecular CH G-bands compared to
their colors. The strength of this molecular feature is correlated with
the carbon abundance, and is measured by the GPE and EGP line indices
defined by \citet{placco2010,placco2011}. Based on the location of a
given star in a GPE vs. EGP diagram, it is possible to infer the level
of its carbon enhancement, regardless of its metallicity. Once
\cemp{} candidates are selected, medium-resolution spectroscopy is
carried out for improved carbon-abundance determinations.

\subsection{Medium-Resolution Spectroscopy}

The stars employed in this work are part of the CEMP candidate list
generated by \citet{placco2011}. Follow-up medium-resolution
spectroscopic observations were carried out in semester 2011B using
the Gemini Multi-Object Spectrograph (GMOS), at the Gemini-S
telescope. The setup included the 600~l~mm$^{\rm{-1}}$ grating in the
blue setting (G5323) and the 1$\farcs$0 slit, covering the wavelength
range of 3300-5500\,{\AA}. This combination yielded a resolving power
of R$\sim2,000$, with an average S/N $\sim40$ at 4300\,{\AA}. The
calibration frames included HgAr and Cu arc lamp exposures (taken
following each science observation), bias frames, and quartz
flats. All tasks related to spectral reduction and calibration were
performed using standard GEMINI/IRAF packages. Table \ref{candlist}
presents details of the medium-resolution observations for each
star.

\begin{figure}[!ht]
\epsscale{1.15}
\plotone{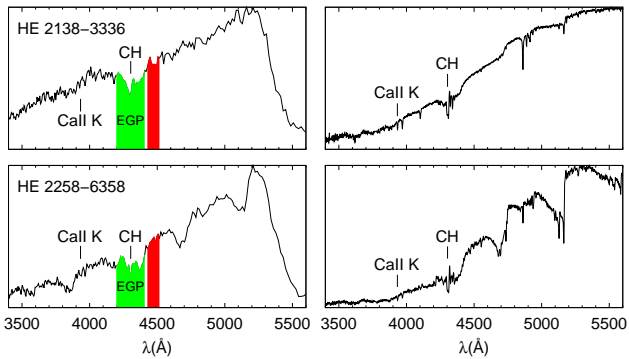}
\caption{Comparison between the low-resolution HES spectra (left panels) and 
GEMINI medium-resolution spectra (right panels).  The green and red bands show,
respectively, the line band and side band of the EGP index, defined
by \citet{placco2011}.}
\label{low_med}
\end{figure}

Figure~\ref{low_med} shows a comparison between the low-resolution
HES and the medium-resolution GMOS spectra of our stars. Even
at low resolution, the CH G-band is clearly seen, and its
strength is captured by the EGP index. In addition, these stars exhibit a
weak Ca\,{\sc{ii}} K line, indicating low metallicity. 
The medium-resolution GMOS spectra are of sufficient quality to
determine estimates of the stellar parameters for the observed stars (see
Section \ref{atmparmed} for further details), and to determine
the metallicity and carbon abundance ratio.

\vspace{0.5cm}

\subsection{High-Resolution Spectroscopy}

The final observational step was to obtain high-resolution spectroscopy
of the most promising targets based on the results of the
medium-resolution spectral analysis. These data were gathered using the
MIKE instrument \citep{mike} on the Magellan-Clay Telescope at Las
Campanas Observatory. We used the 0$\farcs$7 slit with 2$\times$2
on-chip binning, yielding a nominal resolving power of R $\sim35,000$ in
the blue and $\sim28,000$ in the red region, with an average
S/N $\sim85$ at 5200\,{\AA}. MIKE spectra have nearly full optical
wavelength coverage from $\sim$3500-9000\,{\AA}. Table~\ref{candlist}
lists the details of the high-resolution observations for each star.
These data were reduced using a data reduction pipeline developed for
MIKE\footnote{ \href{http://code.obs.carnegiescience.edu/python}
{http://code.obs.carnegiescience.edu/python}.} spectra.

\section{Stellar Parameters}
\label{params}

The stellar parameters (\teff, \logg, \metal) were estimated first from
the medium-resolution spectra, using the procedures described below.
These values were used as first estimates for the determinations based
on the high-resolution spectra.

\subsection{Stellar Parameters from Medium-Resolution Spectra}
\label{atmparmed}

Stellar parameters were determined using the n-SSPP, a modified version
of the SEGUE Stellar Parameter Pipeline \citep[SSPP; see][for a
detailed description of the procedures used]{lee2008a,lee2008b,
allende2008, lee2011, smolin2011}. The n-SSPP is a collection of routines
for the analysis of non-SDSS/SEGUE data that employs both spectroscopic
and photometric ($B_0$, $(B-V)_0$, $(U-B) _0$, $J_0$ and $(J-K)_0$)
information as inputs, to make a series of estimates for each stellar
parameter. Then, using $\chi^2$ minimization in dense grids of synthetic
spectra, and averaging with other techniques as available, the best set
of values is adopted. The internal errors for the stellar parameters
are: 125~$K$ for \teff, 0.25~dex for \logg, and 0.20~dex for \metal.
External errors are of a similar size.

\subsection{Stellar Parameters from MIKE Spectra}
\label{highpar}

The determination of stellar physical parameters from high-resolution
spectroscopy relies on the behavior of the abundances of individual
absorption lines as a function of: (i) the excitation potential, $\chi$, of the
lines from which the abundances are derived (effective temperature);
(ii) the balance between two ionization stages of the same element
(surface gravity) and; (iii) the reduced equivalent width of the lines
measured (microturbulent velocity). The adopted parameters are the
ones that minimize the trend between the line abundances, derived
from the equivalent width of the atomic \ion{Fe}{1} absorption lines,
and the quantities (i), (ii), and (iii). Generally, elemental
abundances are obtained by analysis of both equivalent widths and spectral
synthesis. Equivalent widths are obtained by fitting Gaussian profiles
to the observed atomic lines. For this purpose, we used a line list
based on the compilation of \citet{roederer2010a}, as well as data
retrieved from the VALD database \citep{vald}. Table \ref{eqw} shows
the lines used in this work, with their measured equivalent widths and
abundances.

\begin{figure}[!ht]
\epsscale{1.15}
\plotone{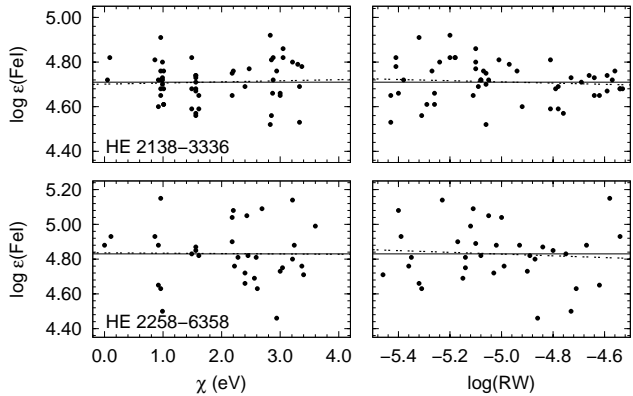}
\caption{\ion{Fe}{1} abundances as a function of the excitation
potential (left panels), and the reduced equivalent width (right
panels). The solid lines mark the average abundance and the dashed
lines represent linear functions fitted to the data.  No clear trends
exist, indicating well-determined estimates of the effective
temperature, $T_{eff}$, and microtubulent velocity, $\xi$.}
\label{feh_comp}
\end{figure}

As seen in Figure~\ref{feh_comp} (left panels), over the range in $\chi$
from 0.0 to $\sim4.5$\,eV, the adopted temperatures ($T_{eff} =$ 5850~K
for \sstar{} and $T_{eff} =$ 4900~K for \rstar) do not present any
significant trend of \ion{Fe}{1} abundances. Likewise, the right panels
of Figure~\ref{feh_comp} show no trends on the behavior of the derived
\ion{Fe}{1} abundances as a function of the reduced equivalent width,
indicating appropriate values for the microturbulent velocity. The same 
applies for estimates of \logg, since the average values of \ion{Fe}{1} and \ion{Fe}{2}
agree within to within 0.02\,dex for both of our stars. 

\begin{figure}[!ht]
\epsscale{1.15}
\plotone{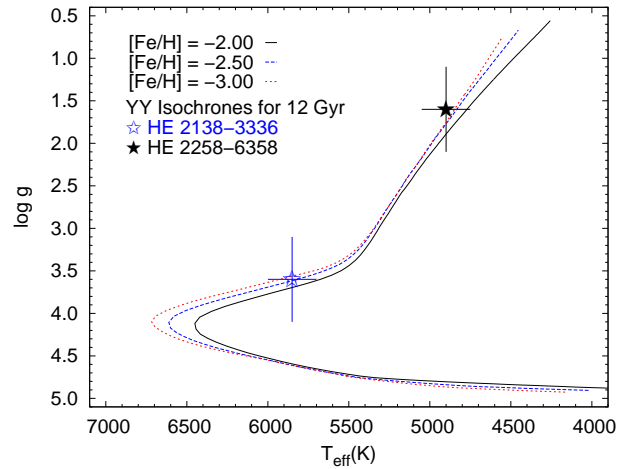}
\caption{The program stars shown in an H-R-Diagram, based on the stellar
  parameters determined from the high-resolution spectra
  (see Table \ref{obstable}). Overplotted are the Yale-Yonsei
  isochrones \citep{demarque2004}, for ages of 12 Gyr and three different values of
  [Fe/H].}
\label{isochrone}
\end{figure}

The final stellar parameters, from both medium- and high-resolution
analysis, are summarized in Table \ref{obstable}. The uncertainties for
the medium-resolution parameters were taken from the n-SSPP, and the
uncertainties for the high-resolution determinations are discussed in
detail in Section \ref{uncertain}. It is worth noting the good agreement
between the \metal{} and \teff{} values for the medium- and
high-resolution spectra. Differences in the \logg{} values arise mainly
due to the difficulty of making this estimate from the medium-resolution
spectra, in particular in the presence of strong carbon features. The
derived effective temperatures and surface gravities from the
high-resolution analysis are shown in Figure~\ref{isochrone}, compared
with 12~Gyr Yale-Yonsei Isochrones \citep{demarque2004} for
\metal=$-$3.0, $-$2.5, and $-$2.0.

\section{Abundance Analysis}
\label{abundsec}

Abundances for individual lines, derived from equivalent widths as well
as from spectral synthesis of some features, are shown in Table
\ref{eqw}. Line abundances obtained through spectral synthesis are 
marked with {\emph{syn}} on the equivalent width column. Our chemical
abundances (or upper limits) for 34 elements, derived from the MIKE
spectra, are presented in Table~\ref{abund1}. A description of our
abundance analysis is given below.
 
\subsection{Techniques}

Our abundance analysis utilizes one-dimensional plane-parallel Kurucz
model atmospheres with no overshooting \citep{castelli2004}. They are
computed under the assumption of local thermodynamic equilibrium (LTE).
We use the 2011 version of the MOOG synthesis code
\citep{sneden1973} for this analysis.  Scattering in this MOOG
version is treated with the implementation of a source function that
sums both absorption and scattering components, rather than treating
continuous scattering as true absorption \citep[see][for further
details]{sobeck2011}. 

Our final abundance ratios, [X/Fe], are given with respect to the solar
abundances of \citet{asplund2009}. Upper limits for elements for which
no absorption lines were detected provide additional information for the
interpretation of the overall abundance pattern of the stars. Based on
the S/N ratio in the spectral region of the line, and employing the
formula given in \citet{frebel2006b}, we derive 3\,$\sigma$ upper
limits for a few elements. All our abundances have been derived with 
LTE models, and where appropriate non-LTE corrections were applied
A summary of the elemental abundances for our
targets is given in Table~\ref{abund1}.

\subsection{Carbon, Nitrogen, and Oxygen}

\begin{figure}[!ht]
\epsscale{1.15}
\plotone{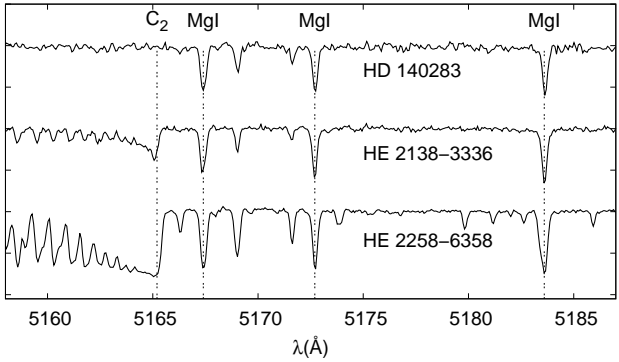}
\caption{Examples of the spectral region around the C$_2$ molecular
  band at 5165\,{\AA} in our stars, compared with HD~140283. Also
  shown is the \ion{Mg}{1} triplet.}
\label{mg_spec}
\end{figure}

\begin{figure}[!ht]
\epsscale{1.15}
\plotone{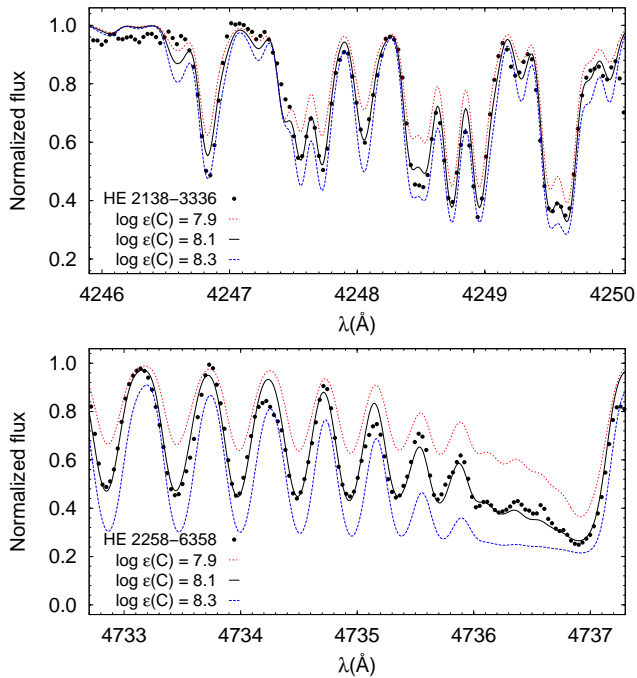}
\caption{Example of a CH band used for carbon-abundance determination
  for \sstar~(upper panel), and a C$_2$ band for \rstar.  
The dots represent the observed spectra, the solid line
  is the best abundance fit, and the dotted and dashed line are the
  lower and upper abundance limits, used to estimate the uncertainty.}
\label{carbon_spec}
\end{figure}

Carbon abundances were derived from both CH ($\lambda$4228\,{\AA},
$\lambda$4230\,{\AA}, and $\lambda$4250\,{\AA}) and C$_2$ ($\lambda$4737
{\AA}, $\lambda$5165\,{\AA}, and $\lambda$5635\,{\AA}) molecular features.
Figure~\ref{mg_spec} shows the C$_2$ band and the \ion{Mg}{1} triplet
for both targets, compared with the spectrum of HD~140283
\citep[\metal=$-$2.2, \teff=5725~K; ][]{sobeck2007}. In carbon-rich
stars, continuum placement is a large source of uncertainty, since the
many strong carbon features compromise its accurate determination.
Figure \ref{carbon_spec} shows two of the features used for the
carbon-abundance determinations for \sstar{} and \rstar. CH $A-X$ band
features are detected between 4240\,{\AA} and 4330\,{\AA} in \sstar, but were
saturated for the cooler \rstar. As seen in Table \ref{eqw}, abundances
derived from CH and C$_2$ features are in good agreement for both stars,
with average values of \cfe=$+$2.43 for \sstar{} and \cfe=$+$2.42 for
\rstar.

\begin{figure}[!ht]
\epsscale{1.15} 
\plotone{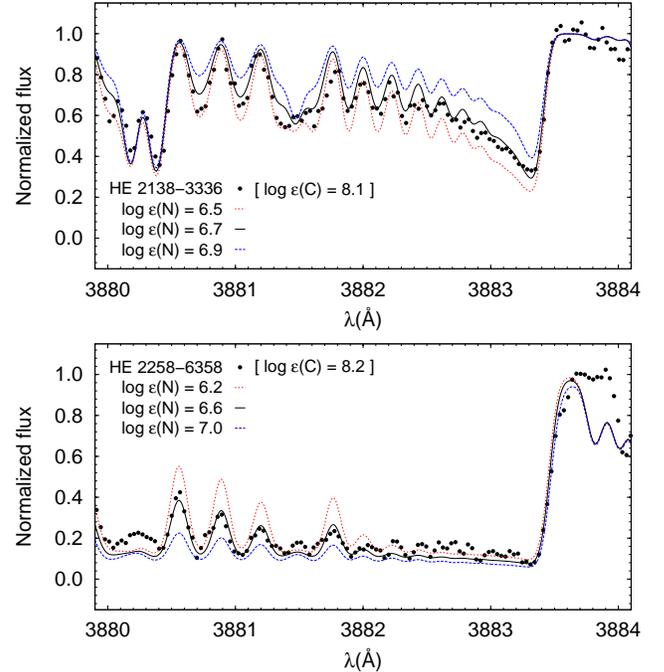}
\caption{Spectral synthesis of the CN band for nitrogen abundance determination.  
  The dots represent the observed
  spectra, the solid line is the best abundance fit, and the dotted
  and dashed line are the lower and upper abundance limits, used to
  estimate the uncertainty.}
\label{nitrogen_spec}
\end{figure}

Nitrogen abundances were determined from spectral synthesis of the CN
band at $\lambda$3883\,{\AA}. For this purpose we used a fixed carbon
abundance, based on the average of the individual abundances determined.
Figure \ref{nitrogen_spec} shows the spectral synthesis for this region
for both targets. In the case of \sstar, the observed spectra agrees
well with the synthetic spectra within 0.2~dex. For \rstar, the band
head appears to be saturated. Even so, it is possible to obtain an
estimate of the nitrogen abundance to within 0.4~dex.

\begin{figure}[!ht]
\epsscale{1.15}
\plotone{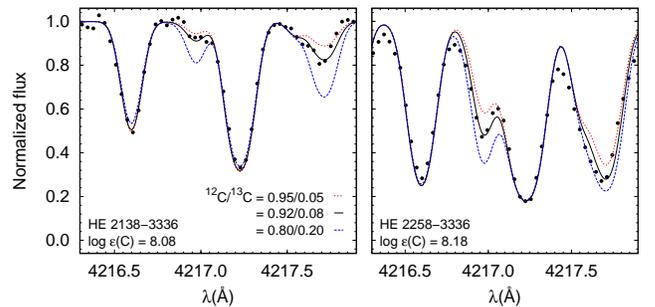}
\caption{Determination of the carbon isotopic ratio $^{12}$C/$^{13}$C
  of our targets.  The dots represent the observed spectra,
  the solid line is the best abundance fit, and the dotted and dashed
  line are the lower and upper abundance limits, used to estimate the
  uncertainty.}
\label{12C13C}
\end{figure}

The $^{12}$C/$^{13}$C isotopic ratio is a sensitive indicator of the
extent of mixing processes in cool red-giant stars. The comparison
between the observed and synthetic spectra used for the determination of
the $^{12}$C/$^{13}$C isotopic ratio is shown in Figure \ref{12C13C}.
Using a fixed elemental carbon abundance, derived from the molecular
features mentioned above, spectra employing three different values of
$^{12}$C/$^{13}$C = 19, 10, and 4 were compared to CH features around
4200\,{\AA}. 

We find that a ratio of about 10 agrees well with the observed spectra 
for both \sstar{} and \rstar. This ratio is consistent with other 
metal-poor CEMP stars \citep{sivarani2006} but is difficult to explain 
with current stellar evolutionary models, which predict much higher ratios for
$^{12}$C/$^{13}$C \citep[e.g., see yields in][]{karakas2010,lugaro2012}.
The low $^{12}$C/$^{13}$C ratios suggest substantial processing of
$^{12}$C into $^{13}$C, where $^{12}$C is accreted from the donor AGB
star and is the dominant isotope produced in the He-burning shells of
the AGB stars. The mechanism for the processing in the donor star is
unknown, although rotational mixing \citep{lagarde2012} and/or
thermohaline mixing \citep{eggleton2008,charbonnel2007,stancliffe2009,
stancliffe2010}, gravity waves \citep{denissenkov2000}, and magnetic
fields \citep{nordhaus2008,busso2007b,palmerini2009} have been proposed
as potential candidates.

Figure~\ref{frebel01} shows the distribution of the carbon and nitrogen
(upper panels) and oxygen (lower panels) abundances for our stars,
compared with literature data collected by \citet{frebel10}. The oxygen
abundance for \rstar{} was determined from the equivalent width of the
$\lambda$6300 {\AA) forbidden line. For \sstar~, no usable oxygen lines were
detected. The carbon and nitrogen abundances in Figure \ref{frebel01}
are in agreement with other stars in the CEMP regime, and the oxygen
abundance for \rstar{} is also among typical values for CEMP stars in
the literature. Interestingly, when comparing the behavior of the
\abund{C}{O} ratio as a function of the carbon abundance for CEMP stars
with \abund{Ba}{Fe} $\geq$ 0.0 (CEMP-s) and \abund{Ba}{Fe} $<$ 0.0
(CEMP-no), one finds that all stars with \abund{C}{O} $>$ 0.0 are
classified as CEMP-s.

\begin{figure}[!ht]
\epsscale{1.15}
\plotone{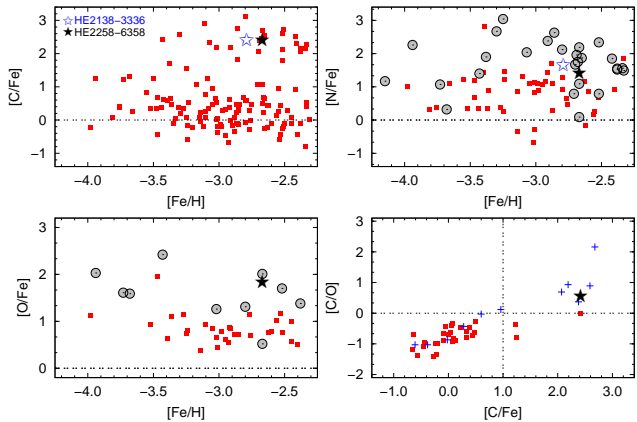}
\caption{Carbon abundances, as a function of the metallicity (upper
panel), for the observed targets, shown as open and filled stars, and
for literature determinations.  The lower panels show \abund{O}{Fe}
vs. \metal~ and \abund{C}{O} vs. \cfe~ for \rstar.  Filled squares
are data collected by \citet{frebel10}
(\metal $< -$2.3), filled circles represent stars from
the same database with \cfe $> +$1.0, and blue crosses show stars with
\abund{Ba}{Fe} $> +$1.0.}
\label{frebel01}
\end{figure}

\subsection{From Na to Zn}

Abundances for Na, Mg, Al, Si, Ca, Sc, Ti, Cr, Mn, Co, and Ni were
determined from equivalent width analysis for both stars, with the
exception of Al and Si for \rstar. For Zn, only upper limits could be
determined. Figure~\ref{frebel02} shows the distribution of the light-
element abundances as a function of metallicity, compared to literature
data \citep{frebel10}. There is no significant difference in the
behavior of the stars from this work and other CEMP stars in this
metallicity range (filled gray circles). This is expected, assuming the
gas which gave birth to the stars was preferentially enriched by massive
SNII \citep{woosley1995}, in addition to the latter pollution by AGB
companions.

\begin{figure*}[!ht]
\epsscale{1.0} 
\plotone{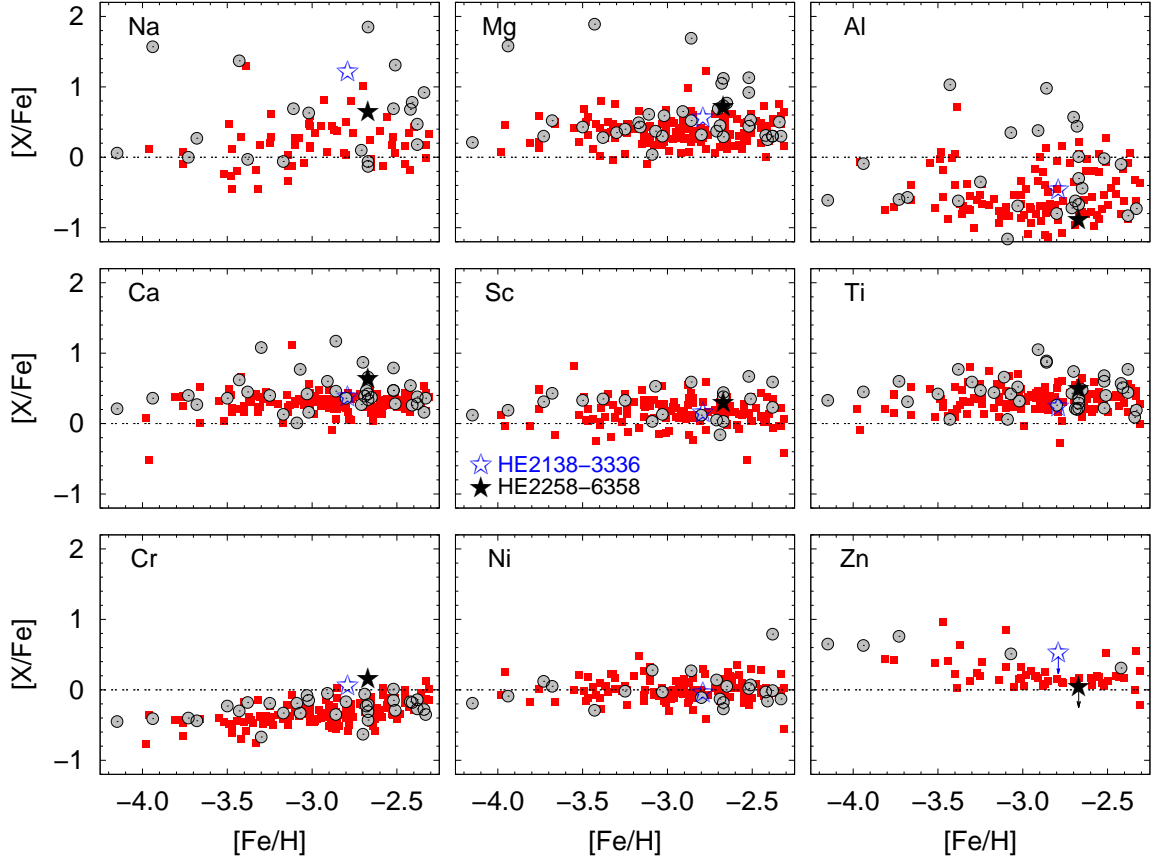}
\caption{Abundance ratios, \abund{X}{Fe}, as a function of 
  metallicity, for selected elements from Na to Zn. Filled squares are
  data collected by \citet{frebel10} (\metal $< -$2.3), and filled
  circles represent stars from the same database with \cfe $> +$1.0.}
\label{frebel02}
\end{figure*}

\subsection{Neutron-Capture Elements}

The chemical abundances for the neutron-capture elements were determined
via spectral synthesis. Figure~\ref{ba_spec} shows the spectra of our
two stars around the $\lambda$4554\,{\AA} Ba line, in comparison with
HD~140283. The synthesis of neutron-capture absorption lines in the blue
spectral region, particularly those close to molecular carbon features,
were often hampered by the strong CH or CN features, and had to be
excluded from the analysis. The results of the abundance determinations
for individual elements and comments on specific features are given
below.

\begin{figure}[!ht]
\epsscale{1.15} 
\plotone{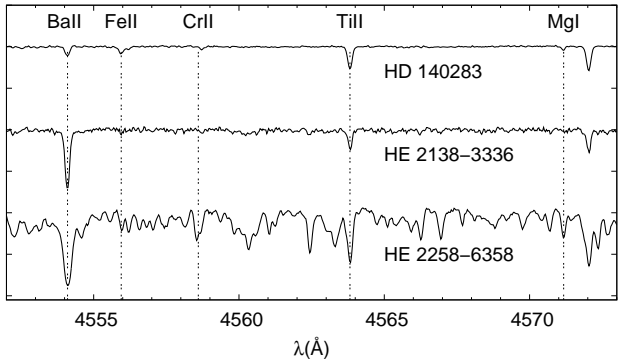}
\caption{Examples of the spectral region around the Ba $\lambda$4554
{\AA} line, compared with HD~140283.}
\label{ba_spec}
\end{figure}

\paragraph{Strontium, Yttrium, Zirconium} 
These three elements belong to the first peak of the $s$-process. Their
abundances are mostly determined from absorption lines in blue
spectral regions, which are affected by the presence of carbon
features.  The Sr $\lambda$4077\,{\AA} and $\lambda$4215\,{\AA} lines were saturated
in the spectrum of \rstar{}, so the abundance was determined from the
$\lambda$4607\,{\AA} line. Three Y lines were found at $\lambda>4800$\,{\AA}
for \rstar, but were not detectable for \sstar, which had its Y
abundance derived from the $\lambda$3774\,{\AA} line. Only one Zr line
($\lambda$4050\,{\AA} for \rstar~ and $\lambda$4208\,{\AA} for \sstar) could be
synthesized for each star. Other Zr features were either too weak or
embedded in carbon molecular bands. The final \abund{X}{Fe} ratios for
Sr, Y, and Zr are slightly overabundant ($>$ +0.3) in both stars, with respect 
to the solar values.

\paragraph{Barium, Lanthanum} 
These elements are representative of the second peak of the $s$-process. 
Ba is strongly overabundant in both
stars. Figure \ref{ba_total} shows the spectral synthesis for the
$\lambda$6496\,{\AA} line. Besides the $\lambda$4554\,{\AA} and
$\lambda$4934\,{\AA} features (saturated for \rstar), additional abundances were calculated
using the $\lambda$5853\,{\AA} and $\lambda$6141\,{\AA}lines. Final abundances are 
\abund{Ba}{Fe}$=$ +1.91 for \sstar~ and \abund{Ba}{Fe}$=$ +2.23 for \rstar. 
Lanthanum is also overabundant in both stars; 
\abund{La}{Fe}$=$ +1.60 for \sstar~ and \abund{La}{Fe}$=$ +1.91 for \rstar. 
There are a number of lines ranging from
4000-6000\,{\AA} suitable for spectral synthesis. Three lines
($\lambda$3995\,{\AA}, $\lambda$4086\,{\AA}, and $\lambda$4123\,{\AA}) were fitted with the
same abundances for \sstar.  Five other features were synthesized for
\rstar, and the values determined from these lines agree within
0.1 dex. 

\begin{figure}[!ht]
\epsscale{1.15} 
\plotone{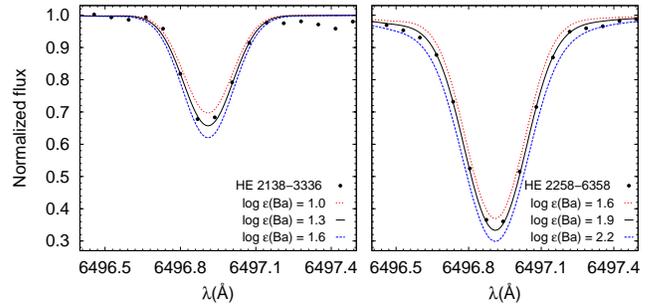}
\caption{Spectral synthesis for the $\lambda$6496\,{\AA} Ba line for \sstar~ and \rstar. 
The dots represent the observed spectra, the solid line is the best
abundance fit, and the dotted and dashed line are the lower and upper
abundance limits, used to estimate the uncertainty.}
\label{ba_total}
\end{figure}

\paragraph{Cerium, Praseodymium, Neodymium, Samarium} 
With the exception of Pr for \sstar, atomic lines for these species
were found in the high-resolution spectra. Cerium and Sm abundances were
determined from lines with $\lambda<4600$\,{\AA}. Even with many lines
available for synthesis, most were weak and blended with carbon
features. Only two Nd features could be synthesized for \sstar, while
9 suitable lines were found for \rstar. Figure~\ref{hf_total} shows, on the
right panels, the spectral synthesis for the $\lambda$4061\,{\AA} Nd line.

\begin{figure}[!ht]
\epsscale{1.15}
\plotone{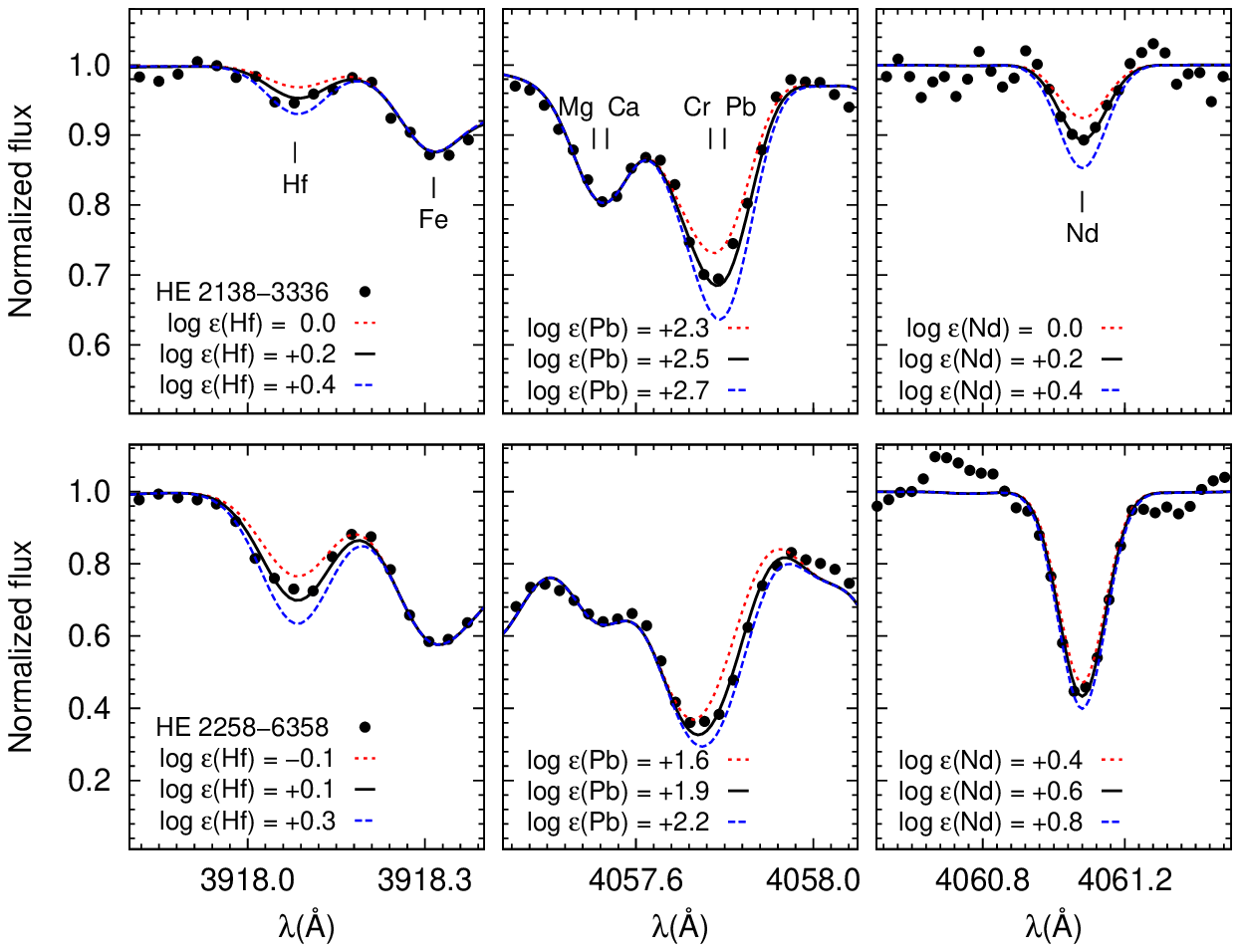}
\caption{Spectral synthesis of selected neutron-capture elements.  The
  dots represent the observed spectra, the solid line is the best
  abundance fit, and the dotted and dashed line are the lower and
  upper abundance limits, used to estimate the uncertainty.}
\label{hf_total}
\end{figure}

\paragraph{Europium} Eu is a well known indicator of $r$-process nucleosynthesis, and its 
abundance helps distinguish the r-only, r/s and s-only abundance regimes
for CEMP stars. There are three lines ($\lambda$3724\,{\AA},
$\lambda$3907\,{\AA} and
$\lambda$6645\,{\AA}) with abundances in reasonable agreement for \rstar~
(average \abund{Eu}{Fe}= +1.68). In the case of \sstar, only an upper
limit could be determined (\abund{Eu}{Fe}$<$ +1.09), based on the
$\lambda$4129\,{\AA} and $\lambda$4205\,{\AA} features.

\paragraph{Gadolinium, Terbium, Dysprosium, Erbium} 
Abundances for these lanthanoids could only be determined for \rstar,
and all features used have $\lambda<4200$\,{\AA}. One Tb feature was
found at $\lambda$3702\,{\AA}, while at least two were found for Gd, Dy and
Er, with the individual line abundances agreeing within 0.2 dex.

\paragraph{Thulium, Ytterbium, Hafnium, Osmium} This set of elements  has many 
suitable features for spectral synthesis at $\lambda<4200$\,{\AA}. No Yb
lines were found for \rstar, while one feature at $\lambda$3694\,{\AA} was
found for \sstar. Figure~\ref{hf_total} shows the spectral synthesis for
the $\lambda$3918\,{\AA} Hf line (left panels).

\paragraph{Lead} This third-peak element is expected to be largely 
produced by the $s$-process \citep{travaglio2001}. Abundances were
determined using two features, $\lambda$3683\,{\AA} and $\lambda$4057
{\AA}. For each star, these two lines yielded the same abundances values
(\abund{Pb}{Fe} = +3.54 for \sstar{} and \abund{Pb}{Fe} = +2.82 for
\rstar). Figure~\ref{hf_total} shows the spectral synthesis for the
$\lambda$4057\,{\AA} line (middle panel). 
If the Pb abundance is determined from its neutral species
(there are only neutral species available in the spectrum),
than the abundance is strongly affected by
non-local thermodynamic equilibrium (NLTE) effects.
Hence, we adopt positive corrections
for Pb abundances of 0.3 dex for \sstar~(\abund{Pb}{Fe}
= +3.84) and 0.5 dex for \rstar~(\abund{Pb}{Fe} = +3.32), 
following \citet{mashonkina2012}. 
We also searched for
possible NLTE corrections for other neutron-capture elements,
such as Sr, Ba \citep{andrievsky2011,bergemann2012} and Eu
\citep{mashonkina2012}. 
However, these measurements are based
on ionized species (e.g. \ion{Sr}{2}, \ion{Ba}{2}, \ion{Eu}{2}). 
In those cases, NLTE effects are expected to be much
smaller than the effects found for neutral species.
Indeed, in all cases, the corrections did not exceed the
quoted uncertainties (0.07-0.15~$dex$) of the abundances, 
so no corrections were applied.

\subsection{Comparisons with Other Very Metal-Poor Stars}

Figure~\ref{frebel03} shows the distribution of the abundances of
selected neutron-capture elements for our program stars as a function
of the metallicity, and compared to the literature data.  Upper limits
on abundances were excluded, and the stars with \cfe$>+$1.0 are marked
as filled circles. No significant differences are found between the
abundances of our targets and the values from literature CEMP-s stars,
for the elements of the first $s$-process peak (Sr, Y, and Zr), as well as
between the second $s$-process (Ba and La) and $r$-process (Eu) peaks
and literature values.

\begin{figure}[!ht]
\epsscale{1.15}
\plotone{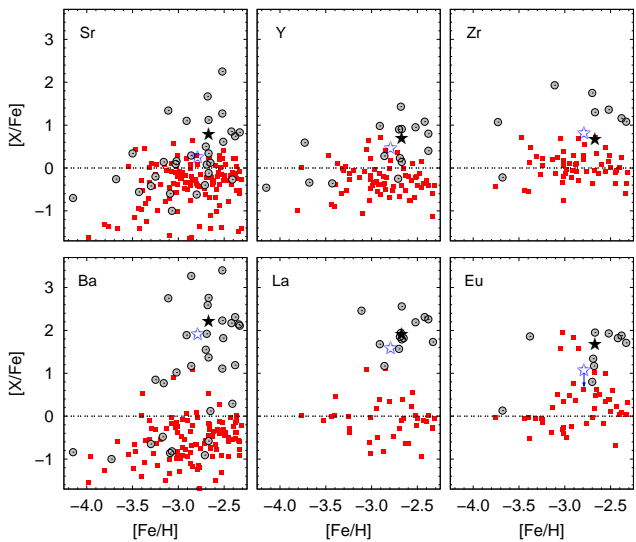}
\caption{Abundance ratios, \abund{X}{Fe}, as a function of 
  metallicity, for selected neutron-capture elements. Filled squares
  are data collected by \citet{frebel10} (\metal $< -$2.3),
  and filled circles represent stars from the same database with
  \cfe $> +$1.0.}
\label{frebel03}
\end{figure}

The differences in the behavior of the abundances of elements formed
by the $s$-process and $r$-process are useful to place constraints on
possible formation scenarios for CEMP stars.  Figure~\ref{frebel04}
presents a [Ba/Fe] vs. [Eu/Fe] diagram for the stars with
\metal $<-$ 2.3 from the literature. Both targets from this work lie in
the same location as other CEMP stars (filled circles). \rstar~ is
close to the limit set for the r/s regime, while the Eu upper limit
for \sstar~ places it in the s-only regime.  More details are provided
in Section \ref{compmod}.

\begin{figure}[!ht]
\epsscale{1.15}
\plotone{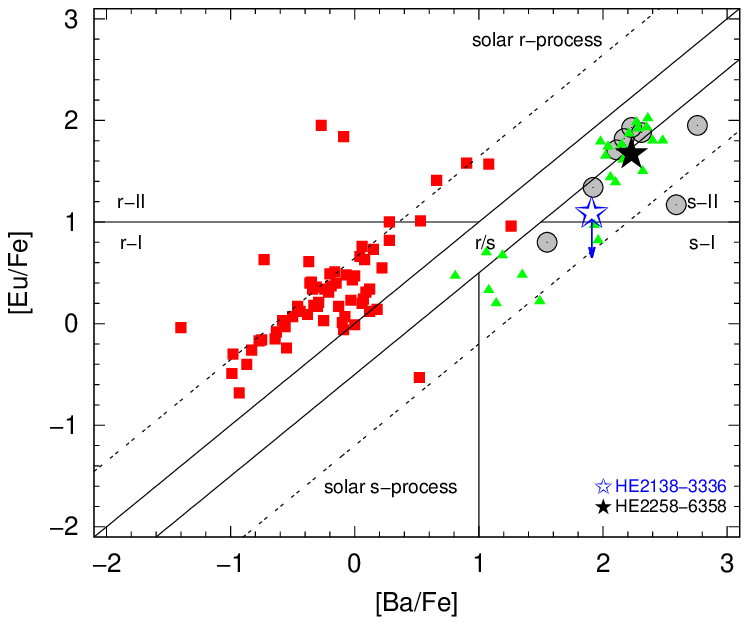}
\caption{\abund{Eu}{Fe} vs. \abund{Ba}{Fe} for the observed targets,
  compared with other neutron-capture-element enhanced stars. The
  solid lines represent the limits for the \cemp~ classes defined by
  \citet{beers2005}, and the dashed lines represent the solar
  $s$-process and $r$-process predictions for the \abund{Ba}{Eu} ratio,
  based on the fractions of \citet{burris2000}.  Filled squares are
  data collected by \citet{frebel10} (\metal $< -$2.3), filled
  circles represent stars from the same database with \cfe $> +$1.0,
  and filled triangles are data from the CEMP stars listed in Table
  \ref{other_stars} (See Section \ref{compmod} for more details).}
\label{frebel04}
\end{figure}

At low metallicity, the $s$-process produces large amounts of Pb, and
high \abund{Pb}{Fe} or \abund{Pb}{Eu} ratios are observational
signatures of the $s$-process operating in metal-poor stars. Based on
models of $s$-process nucleosynthesis in intermediate-mass stars on the
AGB, the minimum $s$-process ratios predicted (\abund{Pb}{Eu} = +0.3) can
be used to set a lower limit on the operation of the $s$-process
\citep{roederer2010b}.  

Figure~\ref{pb} shows the \abund{Pb}{Fe} (left panel) and
\abund{Pb}{Eu} (right panel) ratios of our targets as a function of metallicity.
Also shown are the NLTE corrections explained above. The lead abundances
are within the range presented by other CEMP-s stars. Notably, the value
for \sstar{} (\abund{Pb}{Fe}= +3.54; +3.84 with NLTE correction) is the
highest found to date in metal-poor stars. Also, the \abund{Pb}{Eu}
ratio for both targets (\abund{Pb}{Eu} $=$ +2.45 for \sstar{} and
\abund{Pb}{Eu} = +1.14 for  \rstar) are consistent with the limit set by \citet{roederer2010b}, 
which indicates that the Pb abundances for these stars come from the
$s$-process nucleosynthesis that occured in their AGB companions.

\begin{figure}[!ht]
\epsscale{1.15} 
\plotone{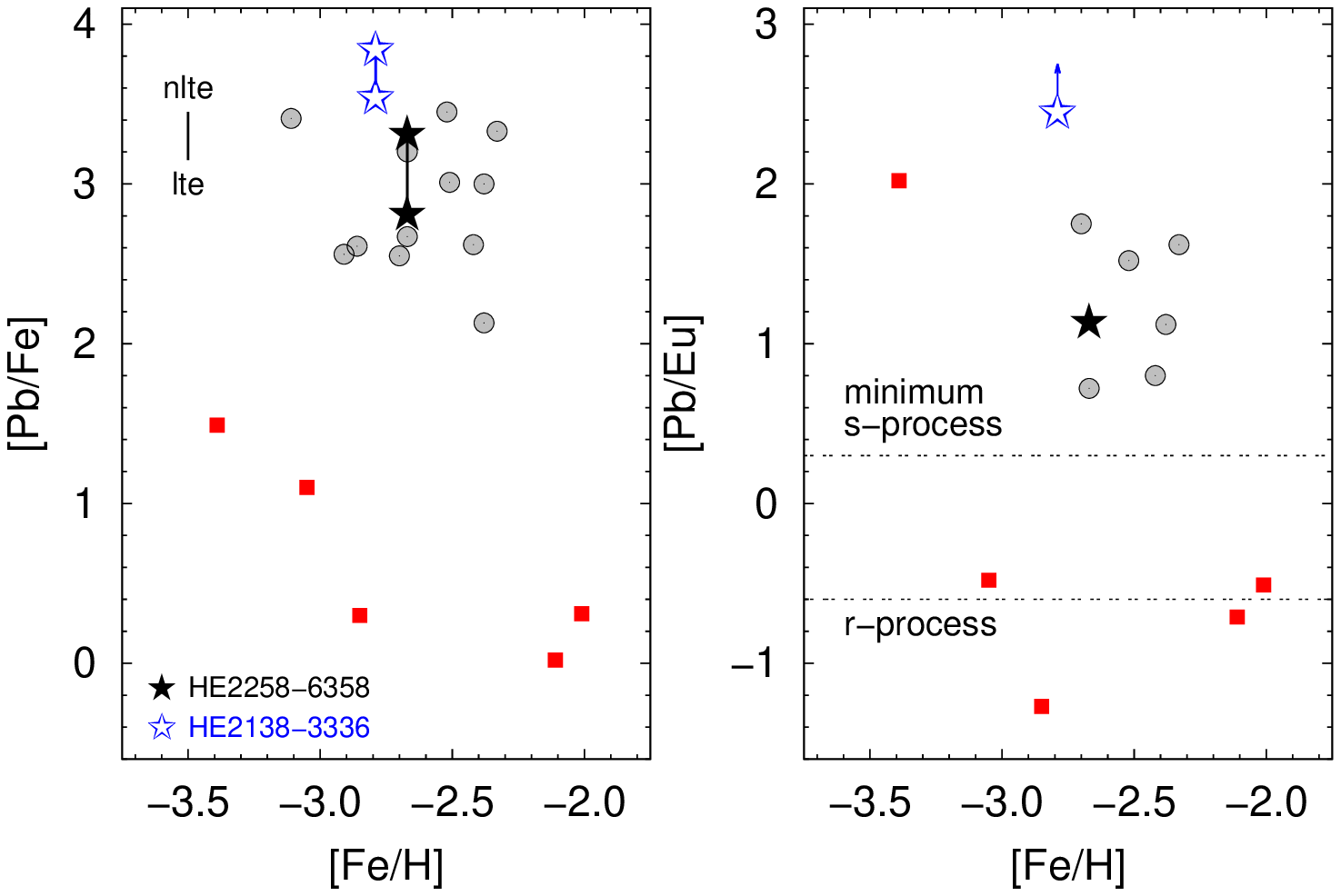}
\caption{\abund{Pb}{Fe} abundance (left panel) and \abund{Pb}{Eu}
  ratio (right panel) for the observed tagets. 
  Filled squares are data from
  \citet{cowan2002,hill2002,sneden2003,masseron2006,ivans2006}, 
  and filled circles represent stars
  with \cfe$>+$1.0 from 
  \citet{aoki2002,lucatello2003,ivans2005,barbuy2005,cohen2006,jonsell2006,aoki2008}.  
 Dashed horizontal lines
  represent limits set by \citet{roederer2010b}.}
\label{pb}
\end{figure}

\vspace{1cm}

\subsection{Uncertainties}
\label{uncertain}

To determine the random uncertainties in our abundance estimates, we
calculate the standard error of the individual line abundances for each
ionization state of each element measured. For any abundance determined
from equivalent-width measurements for less than 10 lines, we determined
an appropriate small-sample adjustment for the standard error
\citep{keeping}. In the case of any abundance uncertainty that was
calculated to be less than the uncertainty in the \ion{Fe}{1} lines, we
adopted the value from \ion{Fe}{1} for that particular element.
Typically, the Fe I standard error is $\sim0.05$ dex. 

For those lines with abundances determined via spectral synthesis,
continuum placement is the greatest source of uncertainty, and it
depends on the S/N of the region containing the particular line. Because of this,
standard errors were determined with two procedures: (i) the uncertainty
was taken directly from the spectral synthesis (e.g., Figure
\ref{hf_total}). Assuming a best value for the abundance of a given
line, lower and upper abundance values were set so they would enclose
the entire spectral feature; (ii) for the elements with two or more
measured abundances, the uncertainties were calculated for small
samples. For those elements with only one line, we adopt 0.1 dex as the
minimum uncertainty for the abundance. Comparing the two values, the larger was
taken as the standard error for the chemical abundances.

To obtain the systematic uncertainties in the abundance estimates, we
redetermined abundances by individually varying the stellar parameters
by their adopted uncertainties. We chose a nominal value of 150\,K
for the uncertainty in the effective temperature, as this value is similar to
the random and systematic uncertainties in the determination of the atmospheric parameters.
The same procedure was also applied for \logg{} (0.5~dex) and $\xi$
(0.3~km/s); results are shown in Table~\ref{syserrstab}. 
Uncertainties in the effective temperature contribute
the most to the abundance uncertainties.  Uncertainties in the surface
gravity are somewhat less important for the abundance determination of
most species. For elements with particularly strong lines, especially
those whose abundances are determined with spectral synthesis, the
microturbulence can be an important source of uncertainty.

\section{Abundance Patterns and Model Comparisons}
\label{discuss}

Our overall aim is to compare the observed elemental abundances of our
CEMP-s stars with the yields from AGB models. This way, constraints on
the mass and metallicity of each progenitor can be obtained, allowing
us to learn about the astrophysical sites of the first/early AGB
nucleosynthesis events and the operation of the $s$-process.

In search of the progenitors of our observed neutron-capture elements
a variety of stars or classes of stars could have been responsible. In
general one first needs to distinguish whether the observed material
was already present in the birth gas cloud or instead reflects a later-time
external enrichment event. Our CEMP-s stars belong to the latter
class, as their strong carbon overabundance in combination with
neutron-capture overabundances associated with the s-process (as
indicated by characteristic abundance ratios, such as Ba/Eu) are a
tell-tale sign of a mass transfer event from a AGB star across a
binary system. 

The detailed study of such individual events (see below) greatly helps
to piece together how the chemical evolution of neutron capture
elements proceeded in the early universe. With few excpetions all
metal-poor stars display some amount of neutron-capture elements in
their surface which can be assumed to reflect the composition of their
birth clouds. At the earliest times, these elements could have
originated from massive, short-lived stars exploding as supernova.
Presumably, some of these supernovae yielded neutron-capture elements
made in the $r$-process. Next-generation stars then formed from
r-process enriched gas. Indeed, most strongly r-process enhanced stars
have a low metallicity of $\mbox{[Fe/H]}\sim-3.0$. In addition, massive
stars that experience strong rotation could have also produced
neutron-capture elements, but through the $s$-process elements. These
stars would, however, produce $s$-process with a different
distribution compared to low-mass AGB stars. Observations of ``normal''
metal-poor stars with $\mbox{[Fe/H]}\sim-3.0$ have help to disentangle
these contributions from the different progenitors.
 
Later in the evolution of the Galaxy, lower-mass, longer-lived AGB
stars began to dominate the production of neutron-capture elements, by
producing $s$-process elements. Accordingly, metal-poor stars born
after the onset of AGB enrichment formed from gas that was
predominently enriched by the s-process. This evolution is somewhat
reflected in the metallicities of metal-poor stars.
\citet{simmerer2004} suggests that the $s$-process may be fully active
at \metal = $-$2.6, but this limit can be as low as \metal = $-$2.8,
according to \citet{sivarani2004} and in individual cases even
lower. Furthermore, \citet{roederer2009b} finds both pure $r$-process
and pure $s$-process enrichment patterns extending over a wide
metallicity range of $-3.0<$~[Fe/H]~$<-0.4$.

Our target stars have \metal$>-$2.6 which canonically places them at a
time when the general s-process production by AGB stars was already
operating. However, in addition to that, our stars show the external
s-process signature that allows us to reconstruct one of those
s-process events occuring in AGB stars. We note here that models for
the s-process in massive stars do not produce large amounts of
Pb. Hence, massive stars cannot be responsible for the abundance
patterns of our CEMP-s stars \citep[e.g., see yields and discussion
by][]{frischknecht2012}. 

We test this whole scenario by first comparing the observed abundances
of our CEMP stars with the scaled Solar System $r$-process pattern
\citep[which is believed to be universal, according to
e.g.,][]{sneden2008}. However, the abundances do not match this
pattern, ruling out an $r$-process origin of the observed
neutron-capture elements. We then compared the observed abundances
with Solar System $s$-process predictions. The abundance do not match
the $s$-process patten across for all elements, especially for the
first-peak elements (Sr, Y, and Zr) as well as Pb. In particular, the
model Pb abundances are underestimated by roughly 1.5 dex. This is not
too surprising because we are working with old, metal-poor stars and
the solar $s$-process pattern represents the integrated yields of AGB
nucleosynthesis over billions of years up to the formation of the Sun.
This natural disagreement, which is found for all CEMP-s and CEMP-r/s
stars, is thought to occur because the $s$-process operates more
efficiently at low metallicities. Owing to the high neutron-to-seed
ratio, this leads to the production of a lot of Pb at early times in
Galactic evolution \citep{gallino1998}. Thus, a comparison of the
observed abundances of our CEMP-s stars with $s$-process model
predictions for low-metallicity AGB stars is required.

In this section we first present the results of new AGB models at the
appropriate metallicities for our targets. Then, a detailed comparison
between the yields of the models and the observed abundance patterns is
presented for each star.

\subsection{AGB Nucleosynthesis Models}
\label{agbmod}

If the observed metal-poor stars have been polluted from material
from a previous AGB companion, then the observed abundances should 
reveal information about the efficiency of mixing events and chemical 
processing that took place during previous evolutionary phases, the 
mass-loss rate during the AGB phase, and the nature of the binary
interaction that took place to pollute the observed star. Furthermore, non-standard
mixing processes, such as thermohaline mixing, may act on the observed
star and alter the accreted composition \citep[e.g.,][]{stancliffe2008,stancliffe2010}.
In light of this complex history, is it possible to explain the observed
abundances using theoretical models of AGB stars? 

Briefly, during the TP-AGB phase the He-burning shell becomes thermally
unstable every $\approx 10^{5}$ years. The energy from the thermal pulse
drives a convective pocket in the He-rich intershell, which mixes the
products of He-nucleosynthesis within this region. The energy provided
by the pulse expands the entire star, pushing the H-shell out to cooler
regions where it is almost extinguished, and subsequently allowing the
convective envelope to move inwards (in mass) to regions previously
mixed by the flash-driven convective pocket. This inward movement of the
convective envelope is known as the third dredge-up (TDU), and is
responsible for enriching the surface in $^{12}$C and other products of
He-burning, as well as heavy elements produced by the $s$-process.
Following the TDU, the star contracts and the H-shell is re-ignited,
providing most of the surface luminosity for the next interpulse period
\citep[see ][ for reviews of AGB evolution and 
nucleosynthesis]{busso1999,herwig2005,straniero2006}.

Details of the TDU phase are notoriously difficult to calculate in
theoretical stellar evolution models \citep[see e.g.,][]{frost1996,
mowlavi1999}. Detailed stellar models suggest that the efficiency or
depth of the TDU increases at low metallicity, which indicates that
low-mass metal-poor AGB stars should be efficient producers of carbon
and $s$-process elements \citep[e.g.,][]{karakas2002,campbell2008,
karakas2010}. Note that theoretical models of stars more massive than
about 3$M_{\odot}$ at metallicities of [Fe/H] $\lesssim -2.3$ show the
signature of hot bottom burning, when the convective envelope is subject
to proton-capture nucleosynthesis via the CN cycle, leading to
nitrogen-rich stars, where [N/C] $> +1$ \citep{johnson2007, pols2012}.
Studies by \citet{izzard2009}, \citet{bisterzo2012}, and
\citet{lugaro2012} have indeed confirmed that many of the CEMP-s stars
should exhibit the signature of low-mass AGB pollution, where the mass
of the polluters is $1.2 \lesssim M / M_{\odot} \lesssim 2.5$. Note that
the latter two studies disagree on the origin of other types of CEMP
stars, such as CEMP-r/s, which show enrichment by both the $r$- and
$s$-process \citep[see][for definitions]{beers2005}. 

The derived elemental abundances for our newly-observed CEMP-s stars
were compared with new predictions from a 1.3$M_{\odot}$ theoretical
model with [Fe/H] = $-2.5$ (with a global metallicity of $Z = 5 \times
10^{-5}$). The AGB evolutionary model was calculated using the Mount
Stromlo Stellar Evolutionary code \citep[][and references
therein]{karakas2010b}, which uses the \citet{vw93} mass-loss rate on
the AGB, and a mixing-length parameter $\alpha = 1.86$. The new AGB
models presented here use updated molecular opacity tables compafred to
those published in \citet{karakas2010b}. We now use the C- and N-rich low
temperature opacity tables from \citet{marigo2009}.

The model was evolved from the zero age
main sequence, through the core helium flash and core helium burning, to
the tip of the AGB. During the AGB, the model experienced 92 thermal
pulses, and evolved to a final core mass of 0.82$M_{\odot}$. 
Out of those 92, 85 have experienced TDU episodes,
with a total of 0.231$M_{\odot}$ of material dredged into the envelope.
This large number of thermal pulses experiencing TDU is quite exceptional 
for a low-mass model, when for example the $M \approx
1.25M_{\odot}$ model of [Fe/H] $= -$2.3 presented in
\citet{karakas2010a} only has 16 thermal pulses. Figure \ref{radius}
shows the evolution of the radius with time for this new AGB model. The
jump in the radius at $2.775\times10^9$~years occurs after the first TDU
event, which causes the star to become carbon rich, and consequently
changes its internal structure. Note that the gap between thermal pulses
shortens considerably.

\begin{figure}[!ht]
\includegraphics[scale=.35,angle=270]{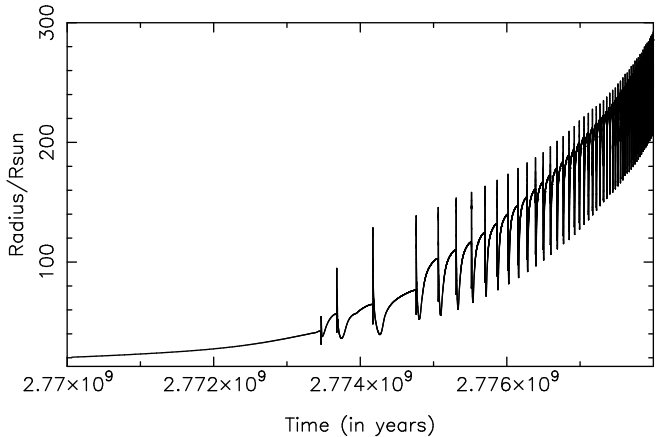}
\caption{Evolution of the stellar radius as a function of time during the TP-AGB 
for the $1.3M_{\odot}$ AGB model presented in this section.}
\label{radius}
\end{figure}

While the behavior of the new 1.3$M_{\odot}$ model during the AGB is unsual 
in that it reaches a high final core mass and experiences many thermal 
pulses, it is consistent and similar to other low-mass models of very low 
metallicity published in the literature. For example, the 0.9 or 1.0$M_{\odot}$ 
models of \metal $\approx -2.3$ \citep[see discussions in][]{karakas2010,lugaro2012}, 
end with relatively high final core masses ($M_{\rm core} >$ 0.7$M_{\odot}$). 
The 0.85$M_{\odot}$ model of $Z=0$ published by \citet{campbell2008} also 
experiences a rapid rate of core growth \citep[as discussed in detail in his PhD thesis;][]{campbell2007}.

The final core mass is set by the competition between mass loss and core growth. 
The rate of core growth observed in the new 1.3$M_{\odot}$ model is rapid toward 
the end of the AGB. This is because 
the H-shell temperature is a function of the chemical composition and the core mass, 
and as the core mass increases so does the H-shell 
temperature and the rate of H-shell burning. A higher core growth rate in 
turn leads to a shortening of the interpulse phase, as the quicker build 
up of H-shell ashes leads to conditions suitable for a TP.

That the core is allowed to reach such high values is mostly as a result of 
the mass-loss rate used on the AGB. We use the \citet{vw93} mass-loss 
prescription, which is a semi-empirical formula derived for stars with 
metallicities of the LMC, SMC, and Galaxy. That is, for stars with much 
higher Z than we are modeling and this adds a significant uncertainty 
into our calculations. The \citet{vw93} formula depend on the radius, 
luminosity and mass so there is an implicit metallicity dependence included.
The new 1.3$M_{\odot}$ model is more compact and consequently has a higher
 effective temperature (by 30\%) compared to the 1.25$M_{\odot}$ model of 
\metal = $-2.3$. 
This is the reason for the large number of TPs. The model star lost 
a total of 0.50$M_{\odot}$ during the AGB (see Figure \ref{radius2}), 
which is only marginally smaller than the amount of mass lost by the 
1.25$M_{\odot}$ of \metal = $-2.3$, which loses about 0.58$M_{\odot}$ 
\citep{karakas2010}.
Furthermore, because the effective temperature never drops 
below 4,000~K that the effect of the updated low-temperature molecular 
opacity tables is minimal \citep[e.g.,][]{marigo2002}.

\begin{figure}[!ht]
\includegraphics[scale=.35,angle=270]{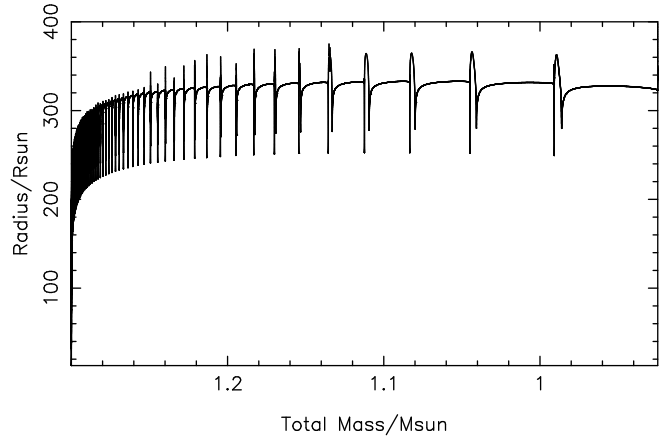}
\caption{Evolution of the stellar radius as a function of mass (in units of $M_{\odot}$)
 during the TP-AGB for the $1.3M_{\odot}$ AGB model presented in this section.}
\label{radius2}
\end{figure}

One of the most significant uncertainties affecting the AGB lifetime of
very low-metallicity AGB models is the interplay between the mass-loss
rate and the surface composition. While these stars have very low iron
abundances, the surface C abundance quickly reaches solar values after
$\approx 10$ thermal pulses, owing to efficient TDU. There is
observational evidence that carbon-enrichment leads to an increase in
the mass-loss rate in cool evolved stars. For example, the near-infrared
colors of carbon stars in the Magallanic Clouds are systematically
redder than those of oxygen-rich stars \citep[e.g., as seen in LMC
surveys such as DENIS and 2MASS][]{cioni2000,nikolaev2000}. At the very
low metallicities of the CEMP stars in the halo there are no direct
observations of the mass-loss process in action. Instead, comparisons
between CEMP stars and model predictions may provide some clues.
Furthermore, it is possible that binary interactions terminate evolution
along the AGB after the star becomes signficantly carbon rich. This
could occur because carbon enrichment leads to an increase in the
opacity, and hence an increase in the radius that causes the primary AGB
star to overflow its Roche lobe. 

The $s$-process abundance predictions were calculated using the
post-processing nucleosynthesis code and full network of 320 species
described in \citet{lugaro2012}, with reaction rates taken from the JINA
REACLIB library \citep{cyburt2010}. For the initial composition we used
the solar distribution of abundances from \citet{asplund2009}, scaled
down to [Fe/H] = $-$2.5. We use the same assumptions about the initial
abundances as outlined in \citet{alves-brito2011}. We did not consider
an initial enhancement for the $\alpha$-elements. The inclusion of an
initial enrichment of $\alpha$-elements does not affect the production
of carbon or neutron-capture elements \citep[e.g.][]{lugaro2012}, which
are the elements we used to determine a good fit between the model AGB
star and the observed metal-poor stars. We also calculate one
nucleosynthesis model where we scaled the iron abundance down to
[Fe/H] = $-$2.8, using the stellar evolutionary sequence described above
as input. This method is not entirely self-consistent, but it provides an
indication of the $s$-process distribution expected at that metallicity
\citep[where a similar method is used by][]{bisterzo2010}.

Low-mass AGB models of very low metallicity can also experience mild
proton-ingestion episodes during the first few TPs and this can shape the
initial s-process distribution \citep[see discussion in][]{lugaro2012}.
We also find proton-ingestion in the 1.3$M_{\odot}$, [Fe/H] = $-2.5$
during the first TP but because the TDU does not begin until much
later, the effect on the final s-process distribution and on the best fit to the
CEMP star abundance distribution is minimal. Note that for the first
$\approx 10$TPs, the $^{13}$C pockets burn under convective conditions
\citep[Regime 3 described in][]{lugaro2012}.

While proton ingestion episodes are found with many evolution codes,
the exact mass and metallicity range over which they occur is not agreed
upon \citep[compare the models of][]{campbell2008,lau2009,suda2010}. 
The issue is further complicated by the fact
that both evolutionary and nucleosynthetic predictions for these events
may be inaccurate as there are many uncertainties associated with them.
The development of hydrodynamical models for this phase of evolution
\citep[e.g.][]{herwig2011,stancliffe2011} may eventually lead to
an improvement in our understanding of such episodes.

In low-mass AGB stars the main neutron source is the $^{13}$C($\alpha$,
n)$^{16}$O reaction, which is activated in the He-intershell region
between thermal pulses at temperatures of $T \gtrsim 90 \times 10^{6}$K.
Observational and theoretical evidence has shown that the
$^{13}$C($\alpha$,n)$^{16}$O reaction is the main neutron source in
low-mass AGB stars of $\approx 1-3M_{\odot}$ \citep{gallino1998,
abia2002}. In order to obtain an enrichment of $s$-process elements, we
artificially introduce some protons into the top of the He-intershell
region. This has become standard practice for such models, for the
simple reason that there is insufficient $^{13}$C in the H-burning ashes
of AGB stars to make it an efficient neutron source. We apply the
assumption that the proton abundance in the intershell decreases
monotonically from the envelope value of $\simeq$ 0.7 to a minimum value
of 10$^{-4}$ at a given point in mass, located at ``M$_{\rm mix}$''
below the base of the envelope \citep{goriely2000}. Protons are
inserted at the deepest extent of each TDU episode, where the
methodology is exactly the same as in e.g., \citet{kamath2012} and
\citet{lugaro2012}, which we refer to for a detailed discussion of the
uncertainties. The mass of the proton profile is a free parameter, which
we set to a constant mass of M$_{\rm mix} = 2 \times 10^{-3} M_{\odot}$.
The protons are captured by the abundant $^{12}$C in the He-intershell
to form a $^{13}$C-rich region in the top 1/10$^{\rm th}$ of the
intershell. Neutrons are then released by the $^{13}$C($\alpha$,n)
$^{16}$O reaction during the interpulse period under mostly radiative
conditions \citep{straniero1995}.

\subsection{Comparison with AGB Models}
\label{compmod}


We compared the observed abundances of our CEMP stars with abundance
predictions from the model described in Section \ref{agbmod}
(1.3$M_{\odot}$ and $Z = 5 \times 10^{-5}$), as well as for 67
different AGB models from \citet{karakas2010} and
\citet{lugaro2012}. These models are in a range of masses of
0.9$M_{\odot}$ to 6.0$M_{\odot}$ and metallicity $Z=10^{-4}$,
including models with $s$- and $r$-process enhancements in their
initial abundance patterns prior to the AGB evolution. The data
include yields for every thermal pulse of each model.

\begin{figure}[!ht]
\epsscale{1.15}
\plotone{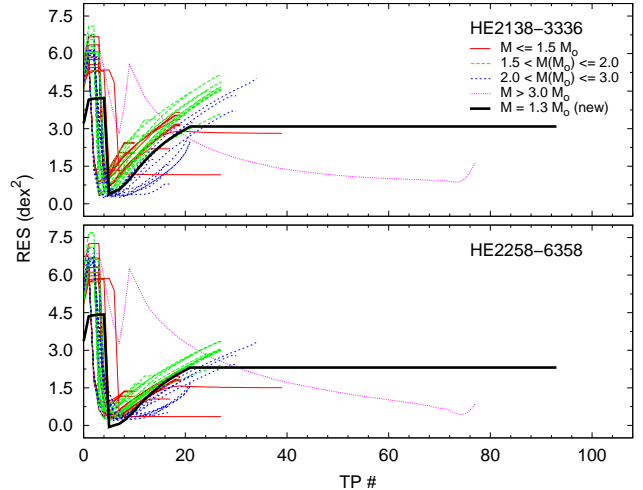}
\caption{Residual values of the model presented in 
Section \ref{agbmod}, and other models from the literature, as a function of 
the thermal pulse number, TP\#.}
\label{res_full}
\end{figure}

To identify the model (and thermal pulse) which best describes the
observed abundances, a residual-like measurement was made. We took the
sum of the squared differences between the abundances of each thermal
pulse of a given model and the observed values, divided by the number of
measured abundances, including all neutron-capture elements and carbon.
Figure \ref{res_full} shows the behavior of this quantity (RES), as a
function of the thermal pulse number, for the model presented in this
work (labeled $new$) and a series of models taken from the literature.
With the exception of models with M $> 3.0~M_{\odot}$, the lowest residual
value is usually reached early in the evolution of the AGB star, for
both \sstar~and \rstar. 

Also in Figure \ref{res_full} it is possible to note that, for \rstar,
the lowest residual value is related to the new low-metallicity model 
described in Section \ref{agbmod}. On the other hand, 
for \sstar, there seems to be a number of
models with masses between 1.5$M_{\odot}$ and 3.0$M_{\odot}$ that yield
lower residual values than the new model. Inspection of the individual
abundances for each of these models reveals that the residuals are
usually lowered by the good agreement for the elements between Ba and
Eu. However, as stated above, those models fail to reproduce the
abundances of the first s-process peak (Sr, Y, and Zr) and Pb.


The interpretation of the abundances in former mass-transfer systems is
complicated by the fate of material accreted from the primary AGB star. Mass
transfer is most likely to occur while the secondary is on the main sequence,
but the secondary can have evolved substantially since then. If the star
develops a deep convective envelope, the accreted material can become
significantly diluted with pristine material from the stellar interior. The
effect of dilution on the surface abundances will then be determined by the
depth of the convective region and the mass of material that was accreted. The
former can easily be supplied by stellar-evolution calculations, but we are
forced to make some assumptions about the latter.

The above picture applies if convection is the only means by which accreted
material can be mixed with the stellar interior. However, accreted material
has undergone nuclear burning, and has a higher mean molecular weight
than the pristine material of the secondary on which it now lies. This
situation is unstable to the process of thermohaline mixing, and the
accreted material can end up being mixed throughout a large portion of
the secondary \citep[see][for example]{stancliffe2007}. The extent of
mixing depends upon the amount of accreted material and also on its
composition. However, the depth to which thermohaline mixing reaches can
be determined by a stellar-evolution code \citep{stancliffe2008}.

We note that, to calculate the residuals, our model abundances were 
not scaled to any of the observed abundance of each star.
This procedure (scaling abundances to, e.g., Ba and Eu) is
commonly used as an attempt to reproduce the relative $s$- and $r$-process
contributions of each element to the observed abundance pattern, and
cannot be used to quantitatively trace the enrichment episodes
experienced by the object. On the contrary, the use of absolute abundances
allows for an assessment of the effects of dilution across the
binary system on the predicted abundance pattern. This can give clues to
differences arising from the interaction between the AGB donor star and the
receiving star. 

We can compute the surface abundance of the secondary for a given
element, $X$, using the equation:

\begin{equation}
X = {M_\mathrm{acc} X_\mathrm{acc} + M_\mathrm{i} X_\mathrm{i} \over M_\mathrm{mix}},
\label{eqdil}
\end{equation}

\noindent where $M_\mathrm{acc}$ is the mass of material accreted, 
$M_\mathrm{mix}$ is the total mass over which material is mixed 
(inlcuding the accreted layer), $M_\mathrm{i}$ is the mass of pristine 
material which will become mixed (i.e., $M_\mathrm{i}  = M_\mathrm{mix} 
- M_\mathrm{acc}$), and $X_\mathrm{acc}$ and $X_\mathrm{i}$ represent 
the accreted and initial compositions of the element. $M_\mathrm{i}$ 
can be determined from stellar models, and depends both on the 
evolutionary state of the object and the nature of the mixing 
mechanisms being considered. 

Typical values for the quantities above were taken from \citet{lugaro2008}. 
We assume the masses of our metal-poor stars to be 0.8~M$_{\odot}$.
When such a star is on its giant branch, its envelope reaches its 
maximum depth, so the outermost 60\% of the star is convective. 
On the other hand, when the star is a subgiant/dwarf, 
this depth can be as low as 5\%. The free parameters 
are $M_{acc}$, and the details of the AGB evolution, nucleosynthesis and
mass transfer. \citet{lugaro2008} find that, for their particular case, the 
currently observed fluorine-rich CEMP-s star\footnote{Fluorine can be obtained 
with infrared spectra, hence its abundance could not be
determined for the CEMP stars presented in this work.}
should have accreted
between 0.05~M$_{\odot}$ and 0.12~M$_{\odot}$ from its AGB companion, so
a value of 0.10~M$_{\odot}$ was adopted in our case.
In addition, the initial abundance
pattern of the receiver star is taken to be the same as that of the
donor prior to its AGB evolution, which is the solar distribution of
abundances from \citet{asplund2009}, scaled down to \metal=$-$2.5.

It should be stressed that the assumption of 0.1~M$_{\odot}$ of accreted
material may not be truly representative. The fluorine-rich star discussed in 
\citet{lugaro2008} is an unusual object whose high level of fluorine 
enrichment requires a large mass of AGB ejecta to have been accreted 
\citep[see also][]{stancliffe2009}. In fact,  the population synthesis 
modeling of \citet{izzard2009} suggests that many systems accrete very
little material. However, these models are based upon the use of a
Bondi-Hoyle prescription for wind accretion, and a better treatment of
the wind may lead to more accretion; it may be possible to accrete as
much as 0.4 M$_\odot$ in exceptional circumstances \citep[see][for
details]{abate2013}.



In applying the above equation, we must take care that we correctly
identify the depth of mixing. If we include only the action of convection,
then when the star reaches the subgiant branch its convective envelope is not
yet deep enough, and still lies within the accreted layer. No dilution of
accreted material will have taken place so far. However, if thermohaline
mixing is taken into account, accreted material will have mixed to a depth of
around 0.5 M$_\odot$ from the surface \citep{stancliffe2008}. Thus, we
distinguish two cases for \sstar~ (\logg = 3.6): one with no dilution
of accreted material, and one where we use $M_\mathrm{mix}  = 0.5$ 
and $M_\mathrm{i} = 0.4$ in our dilution equation.

\begin{figure}[!ht]
\epsscale{1.15}
\plotone{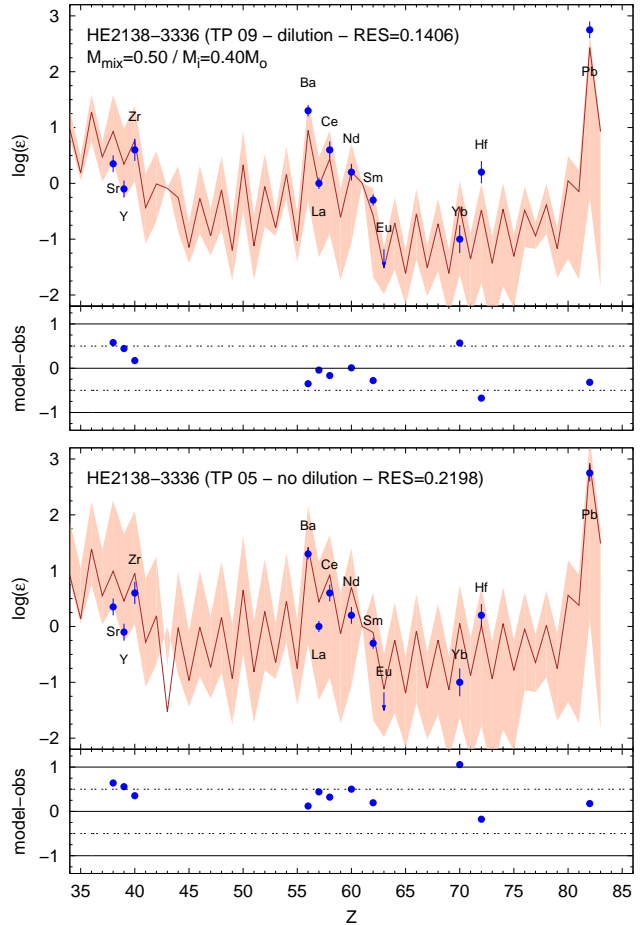}
\caption{Abundance pattern of \sstar, compared with the $M_\mathrm{mix}  = 0.5$ 
M$_\odot$ dilution case (top) and the non-dilution case (bottom). The model used
is the one described in Section \ref{agbmod}.
The shaded area covers the model prediction ranges from the
initial to the final abundances, and the solid line shows the abundance
pattern for the thermal pulse having the lowest residual.
The residual is calculated by the
sum of the squared differences between the abundances of the thermal
pulse and the observed values, divided by the number of measured abundances.}
\label{abpat21b}
\end{figure}

Figure \ref{abpat21b} shows the abundance pattern of \sstar~in two
cases. The top panel compares the observed abundances with the model
described in Section \ref{agbmod}, including dilution according to
Equation \ref{eqdil}; this case corresponds to the dilution of material
by the action of thermohaline mixing. The lower panel shows the same
comparison without dilution, i.e., for the case where only standard
convection is considered. The shaded area covers the model prediction
ranges from the initial to the final abundances, and the solid line
shows the abundance pattern for the thermal pulse having the lowest
residual. Indicated in each panel are the number of the best-matching
thermal pulse and the value of the residual. It is worth noticing the
agreement (within $\sim2\sigma$) between the observed abundance pattern
and the model including dilution. In this case, the values agree within
around 0.5 dex, with the exception of Hf. For the model without
dilution, nearly all the residuals are positive, suggesting an
overproduction of all the heavy elements. 



Figure \ref{res_plot} shows the residuals calculated for each thermal
pulse, including dilution, compared with the residuals for the case without
dilution included. For \sstar, the best residual is at TP\#9 in the
dilution case, whereas it is at TP\#5 for the non-dilution case. The
residual is significantly lower for the case with dilution. 

\begin{figure}[!ht]
\epsscale{1.15}
\plotone{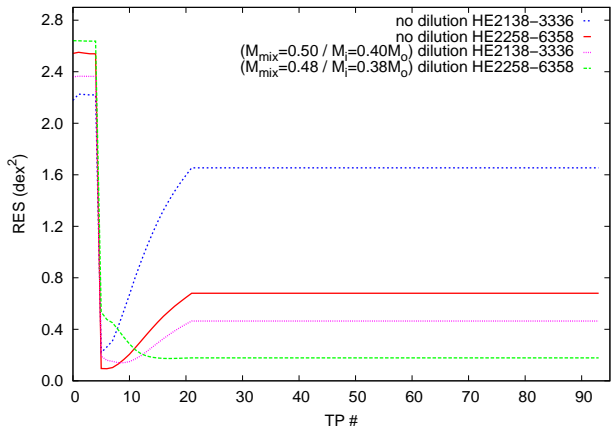}
\caption{Residual values, as a function of the thermal pulse number, for
\sstar~and \rstar, with and without dilution.}
\label{res_plot}
\end{figure}

The fact that our best fit, including dilution, happens at a relatively
early pulse may point to the fact that the mass transfer in this system
took place before the donor star was able to complete its full AGB
evolution. Note that the AGB model used here enriches itself with carbon
and s-elements between pulses 5 to 21. 
There are two possible reasons for the AGB star to transition to either 
a post-AGB star or directly into a white dwarf -- either the mass transfer 
happened earlier as a result of binary interaction, or the mass-loss 
rates used in the model are not high enough.
The fact that radial velocity variation is detected
in this object may tentatively support the former hypothesis, as this
suggests this system may be a close binary. However, we cannot rule out
the possibility that the heavy element distribution could be better fit
by a different mass of AGB donor, or if the s-process distribution is
affected by uncertainties in the $^{13}$C pocket \citep[see
e.g.][]{bisterzo2010,lugaro2012}.



For the case of HE~2258-6358 (log $g = 1.8$), the situation is somewhat
simpler, because the star has undergone first dredge-up, which takes
place between $2.3 < \log g < 3.2$ \citep[e.g.][]{stancliffe2009b}. Any
accreted material must have become mixed by convection at this point,
even if it was unaffected by any other physical process while on the
main sequence. The convective envelope reaches a depth of around 0.48
M$_\odot$. This is comparable to the depth of thermohaline mixing, so we
can examine the two cases with a single calculation, namely
$M_\mathrm{mix} = 0.48 M_\odot$, $M_\mathrm{acc} = 0.1 M_\odot$ and
$M_\mathrm{i} = 0.38 M_\odot$.

The abundance pattern of \rstar~is shown in Figure \ref{abpat22b}. The
best residual behavior occurs from TP\#17 to the final abundance,
assuming a mixing depth of $M_\mathrm{mix} = 0.48$ M$_\odot$. However,
in this case the residual value is worse than in the case of
non-dilution (see Figure~\ref{res_plot}), with the non-dilution case
having its best fit at TP\#6. This result is problematic because \rstar~
is a post-first dredge-up object. Any accreted material would have been
mixed by the deepening of the convective envelope, if it had not
previously been mixed by some other non-convective process. Although the
dilution case yields higher residuals than the case with no dilution
included, this behavior could be due to the possible r/s origin of
\rstar, as seen in Figure \ref{frebel04}. The observed values, especially
for the elements from the second $r$-process peak, are consistently
above the dilution model values. However, this does not apply for the Pb
abundance, which is mainly formed by the $s$-process. This could be due to an
underestimated initial abundances of the receiver star, which may have
formed from a previously $r$-process enriched cloud. This previous
enrichment could account for the differences between the observed
abundances and the model values.

\begin{figure}[!ht]
\epsscale{1.15}
\plotone{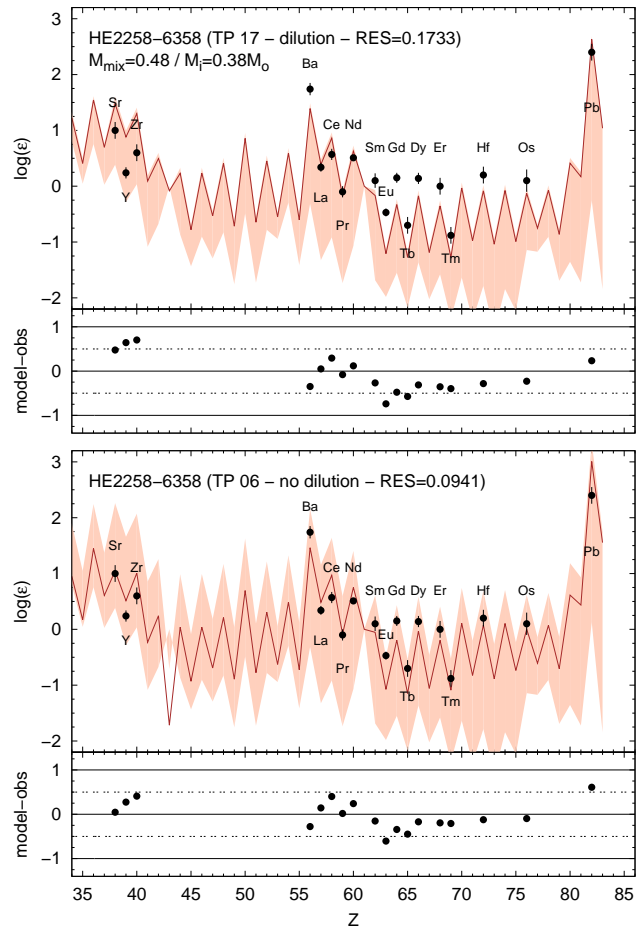}
\caption{Abundance pattern of \rstar, compared with the 60\% depth
dilution case (top) and the non-dilution case (bottom). The
lines and symbols are the same as Figure \ref{abpat21b}.}
\label{abpat22b}
\end{figure}

\citet{jonsell2006} lists several
scenarios for CEMP-r/s production, including a binary system formed from
an $r$-process-rich interstellar medium along with AGB pollution. They
argue that the probability of forming a star from a pre-enriched
$r$-process cloud, independent of AGB enrichment from a companion, is
negligible, and suggest that the formation of the binary could be
triggered by a supernova capable of producing $r$-process elements. In
any case, in order to support this hypothesis, the nature of the binary
(assuming it is such) for \rstar~must be determined by radial velocity
monitoring.



To investigate this issue further, 
this analysis was extended to known CEMP-s and CEMP-r/s
stars from the literature, and results are presented in Table
\ref{other_stars}. The stars selected have at least 7 abundances
determined, including neutron-capture elements (up to 21 different
species) and carbon. For the 26 stars in the table, the residuals were
calculated for two cases: no dilution of accreted material, and the
dilution over a mass range of 0.48 M$_\odot$ (corresponding to the case
of dilution by either the convective envelope in the case of giants and
by thermohaline mixing in the case of less evolved objects). The
assumption of dilution does indeed present better residuals for
\logg $<$ 2.5, while the no dilution case tends to give better fits in the
case of objects with higher \logg. In the cases where no dilution gives
the best fit, just over half the objects have their lowest residual at
higher pulse numbers. For the dilution cases, three quarters of them
have the their lowest residuals for pulse numbers of 10 or lower. 

The reasons why low surface gravity objects can be fit by
neutron-capture abundances of AGB stars having undergone few thermal 
pulses is not easy to explain. 
An early truncation of the AGB phase due to
the presence of a companion should be rare, as few CEMP stars are found
in close binary systems \citep{lucatello2003}. In addition, the fate of
the AGB donor should be independent of the present-day state of CEMP
star, as mass transfer likely happened many gigayears ago. Clearly there
is still further work needed in order to understand the nature of these
systems. One note of caution should be added.  We have only tried
to fit the abundances from a single AGB star, and we have considered only
one mass width for the partially-mixed zone that gives rise to the
$^{13}$C pocket in the AGB nucleosynthesis calculation. 
It remains to be seen whether other choices of these
parameters would result in improved fits.

Finally, by comparing the abundance patterns of \rstar{} and the
CEMP-r/s stars listed in Table \ref{other_stars} (green filled triangles
in Figure \ref{frebel04}), it is possible to note that all the CEMP-r/s
analyzed show the same overproduction of the second peak $r$-process
elements when compared with the AGB model presented in this work.
This behavior is in agreement with recent studies \citep{bisterzo2012,lugaro2012},
which suggest that one possible scenario for the formation of the 
CEMP-r/s is a binary system formed in a molecular cloud pre-enriched
with $r$-process elements.
Moreover, 12 out of the 13 r/s stars in Table \ref{other_stars} present
their lowest residual value for the non-dilution case, similar to the
behavior of \rstar.

\section{Conclusions}
\label{final}

In this paper we have analyzed two newly-discovered CEMP stars, and
compared their abundance patterns with yields from a low-mass,
metal-poor AGB model, including the effects of dilution across a binary
system. 

Our targets were initially selected due to the presence of characteristic
carbon-enhancement features in low- and medium-resolution spectra
\citep{placco2010,placco2011}. The high-resolution follow-up spectra
reported here allowed the determination of abundances (or upper limits)
for 34 elements; light elements ($Z<30$) are in agreement with those of other 
typical halo stars, but their neutron-capture element abundances indicate 
these stars should be CEMP-s stars, enriched by the $s$-process.
One of our targets, \sstar,
exhibits a remarkably high Pb abundance ratio, supporting the hypothesis of
relative lead overproduction in AGB stars at low metallicities. Both stars exhibit
$s$-process signatures in their abundance patterns at metallicities
\metal = $-$2.7 and $-$2.8, agreeing with statements that the onset of the
s-process can be as early as \metal = $-$2.8 \citep{sivarani2004}. We also
confirm the CEMP-s classification of \sstar, through its abundance
pattern and observed changes in radial velocity over the period of one
year, and argue for the possible CEMP-r/s classification of \rstar,
because no $s$-process model agrees well with the observed abundance pattern, 
likely due to a contribution of these elements by the $r$-process.

We compared our derived abundances with low-metallicity $s$-process models, 
assuming mass transfer across a binary system. We also took into account the
effects of dilution in the binary system.
The use of non-scaled abundances (e.g. with respect to Ba and Eu), 
when comparing the observations
with yields from AGB models, 
opens a new opportunity for determining
the characteristics of the progenitors of CEMP-s stars, allowing the
study of dilution effects on the surface abundances of the AGB star. Extension
of this analysis to other CEMP-s and CEMP-r/s stars found in the
literature shows how complex the interaction in a binary system can be,
and places constraints on the timescales and conditions which allow the
mass transfer to take place, and generate the abundance pattern observed
today.

\acknowledgments V.M.P. and S.R. acknowledge FAPESP 
(2010/{}08996-0 and 2012/{}13722-1) 
and, INCT-A funding.  T.C.B. acknowledges partial support
from grant PHY 08-22648; Physics
Frontier Center/Joint Institute or Nuclear Astrophysics (JINA),
awarded by the US National Science Foundation. 
A.I.K. acknowledges support for this work through an ARC Future Fellowship 
(FT110100475), and the NCI National Facility at the ANU.
N.C. acknowledges
support by Sonderforschungsbereich SFB 881 ``The Milky Way System''
(subproject A4) of the German Research Foundation (DFG).  
R.J.S. is the recipient of a Sofja Kovalevskaja 
Award from the Alexander von Humboldt Foundation. These
results are based on observations obtained at the Gemini South
(GS-2011B-Q-91). Gemini Observatory is operated by the
Association of Universities for Research in Astronomy, Inc., under a
cooperative agreement with the NSF on behalf the Gemini partnership:
the National Science Foundation (United States), the Science and
Technology Facilities Council (United Kingdom), the National Research
Council (Canada), CONICYT (Chile), the Australian Research Council
(Australia), CNPq (Brazil), and CONICET (Argentina).

\clearpage

\begin{center}
\begin{deluxetable}{lrr}
\tabletypesize{\scriptsize}
\tablecaption{Observationally Data for the Observed Candidates \label{candlist}}
\tablehead{
\colhead{} & 
\colhead{HE~2138$-$3336} &
\colhead{HE~2258$-$6358}}
\tablewidth{300pt}
\startdata
$\alpha$ (J2000)  &    21:41:20.4 &       23:01:48.6 \\
$\delta$ (J2000)  & $-$33:22:29.0 &    $-$63:42:24.0 \\
V (mag)           &          15.0 &             14.5 \\
\jk               &          0.43 &             0.58 \\
GPE (\,{\AA})     &          41.7 &             56.9 \\
EGP (mag)         &       $-$0.49 &          $-$0.28 \\
\hline
\multicolumn{3}{c}{Medium Resolution -- Gemini/GMOS} \\
\hline
Date              &    2011 07 20 &       2011 09 03 \\
UT                &      08:25:40 &         07:03:49 \\
Exptime (s)       &           800 &              960 \\
v$_{r}$(km/s)     &          63.9 &            103.6 \\
\hline
\multicolumn{3}{c}{High Resolution -- Magellan/MIKE} \\
\hline
Date              &    2011 11 03 &       2011 11 03 \\
UT                &      02:55:33 &         00:44:53 \\
Exptime (s)       &          3000 &             1800 \\
v$_{r}$(km/s)     &          56.1 &            102.1 \\
S/N (4000\,{\AA}) &            40 &               32 \\
S/N (4500\,{\AA}) &            58 &               47 \\
S/N (5200\,{\AA}) &            60 &              100 \\
\hline
Date              &    2012 09 09 &          \nodata \\
UT                &      07:13:14 &          \nodata \\
Exptime (s)       &           860 &          \nodata \\
v$_{r}$(km/s)     &          29.8 &          \nodata
\enddata
\end{deluxetable}
\end{center}

\begin{center}
\begin{deluxetable}{ccccccccc}
\tabletypesize{\scriptsize}
\tablecolumns{9}
\tablewidth{400pt}
\tablecaption{\label{obstable} Derived Stellar Parameters}
\tablehead{
&
\multicolumn{3}{c}{Medium Resolution}&&
\multicolumn{4}{c}{High Resolution}\\
\cline{2-4} \cline{6-9}\\ 
\colhead{}&
\colhead{\teff{}(K)    }&
\colhead{\logg{}(cgs)  }&
\colhead{\metal        }&
\colhead{              }&
\colhead{\teff{}(K)    }&
\colhead{\logg{}(cgs)  }&
\colhead{$\xi$(km/s)   }&
\colhead{\metal{}      }}
\startdata
HE~2138$-$3336 & 6036 (125) & 2.62 (0.25) & $-$2.53 (0.20) && 5850 (150) & 3.60 (0.50) & 1.60 (0.30) & $-$2.79 (0.01) \\
HE~2258$-$6358 & 4753 (125) & 0.81 (0.25) & $-$2.94 (0.20) && 4900 (150) & 1.60 (0.50) & 2.00 (0.30) & $-$2.67 (0.03)
\enddata
\end{deluxetable}
\end{center}

\ltab

\begin{deluxetable}{lrrrrrrr} 
\tabletypesize{\tiny}
\tablecolumns{5} 
\tablewidth{0pc} 
\tablecaption{\label{eqw} Equivalent Width Measurements} 
\tablehead{
\colhead{}&
\colhead{}&
\colhead{}&
\colhead{}&
\multicolumn{2}{c}{\sstar}&
\multicolumn{2}{c}{\rstar}\\
\cline{5-6} \cline{7-8}\\ 
\colhead{Ion}& 
\colhead{$\lambda$}& 
\colhead{$\chi$} &  
\colhead{$\log\,gf$}& 
\colhead{$W$}&
\colhead{$\log\epsilon$\,(X)}&
\colhead{$W$}&
\colhead{$\log\epsilon$\,(X)}\\
\colhead{}& 
\colhead{({\AA})}& 
\colhead{(eV)} &  
\colhead{}& 
\colhead{(m{\AA})}&
\colhead{}&
\colhead{(m{\AA})}&
\colhead{}} 
\startdata  
C    CH   & 4228.00 &  \nodata & \nodata & syn &  8.1 & \nodata & \nodata \\ 
C    CH   & 4230.00 &  \nodata & \nodata & syn &  8.1 & \nodata & \nodata \\ 
C    CH   & 4250.00 &  \nodata & \nodata & syn &  8.1 & syn &  8.0 \\ 
C    C2   & 4737.00 &  \nodata & \nodata & \nodata & \nodata & syn &  8.1 \\ 
C    C2   & 5165.00 &  \nodata & \nodata & \nodata & \nodata & syn &  8.4 \\ 
C    C2   & 5635.00 &  \nodata & \nodata & \nodata & \nodata & syn &  8.2 \\
N    CN   & 3883.00 &  \nodata & \nodata & syn & 6.7 & syn &  6.6 \\
O I   & 6300.30 & 0.00 & $-$9.82 & \nodata & \nodata & 32.4 & 7.9 \\
Na I  & 5889.95 & 0.00 &    0.11 & 194.0 &   4.8 & 192.8 &   4.2 \\
Na I  & 5895.92 & 0.00 & $-$0.19 & 143.9 &   4.6 & 169.2 &   4.2 \\
Mg I  & 4702.99 & 4.33 & $-$0.38 &  49.2 &   5.3 &  \nodata &  \nodata \\
Mg I  & 5172.68 & 2.71 & $-$0.45 & 160.7 &   5.4 & 202.7 &   5.4 \\
Mg I  & 5183.60 & 2.72 & $-$0.24 &  \nodata &  \nodata & 326.7 &   5.9 \\
Mg I  & 5528.40 & 4.34 & $-$0.50 &  47.1 &   5.4 &  \nodata &  \nodata \\
Al I  & 3961.52 & 0.01 & $-$0.34 &  87.7 &   3.2 &  syn &  2.9 \\
Si I  & 4102.94 & 1.90 & $-$3.14 & \nodata & \nodata & syn & 5.2 \\
Ca I  & 4454.78 & 1.90 &    0.26 &  55.3 &   4.1 &  \nodata &  \nodata \\
Ca I  & 5588.76 & 2.52 &    0.21 &  17.8 &   3.9 &  59.8 &   4.2 \\
Ca I  & 5594.47 & 2.52 &    0.10 &  \nodata &  \nodata &  69.7 &   4.5 \\
Ca I  & 6162.17 & 1.90 & $-$0.09 &  35.8 &   4.0 &  89.1 &   4.3 \\
Ca I  & 6439.07 & 2.52 &    0.47 &  21.8 &   3.8 &  \nodata &  \nodata \\
Sc II & 4246.82 & 0.32 &    0.24 &  syn  &   0.2 &  \nodata &  \nodata \\
Sc II & 5641.00 & 1.50 & $-$1.13 &  \nodata &  \nodata &  10.9 &   0.8 \\
Sc II & 5657.91 & 1.51 & $-$0.60 &  \nodata &  \nodata &  26.3 &   0.8 \\
Ti I  & 5210.39 & 0.05 & $-$0.83 &  \nodata &  \nodata &  61.5 &   3.0 \\
Ti II & 3759.29 & 0.61 &    0.28 &  98.5 &   2.3 &  \nodata &  \nodata \\
Ti II & 3913.46 & 1.12 & $-$0.42 &  55.7 &   2.4 &  \nodata &  \nodata \\
Ti II & 4443.80 & 1.08 & $-$0.72 &  43.7 &   2.3 &  \nodata &  \nodata \\
Ti II & 4468.52 & 1.13 & $-$0.60 &  53.5 &   2.5 &  \nodata &  \nodata \\
Ti II & 4501.27 & 1.12 & $-$0.77 &  36.8 &   2.3 &  \nodata &  \nodata \\
Ti II & 4533.96 & 1.24 & $-$0.53 &  52.0 &   2.5 &  \nodata &  \nodata \\
Ti II & 4563.77 & 1.22 & $-$0.96 &  37.1 &   2.6 &  \nodata &  \nodata \\
Ti II & 4571.97 & 1.57 & $-$0.32 &  46.1 &   2.5 &  \nodata &  \nodata \\
Ti II & 5185.90 & 1.89 & $-$1.49 &  \nodata &  \nodata &  38.7 &   2.7 \\
Ti II & 5226.54 & 1.57 & $-$1.26 &  \nodata &  \nodata &  87.6 &   3.0 \\
Ti II & 5381.02 & 1.57 & $-$1.92 &  \nodata &  \nodata &  43.1 &   2.9 \\
Cr I  & 4254.33 & 0.00 & $-$0.11 &  67.2 &   2.8 &  \nodata &  \nodata \\
Cr I  & 5206.04 & 0.94 &    0.02 &  23.8 &   2.6 &  95.3 &   3.0 \\
Cr I  & 5208.42 & 0.94 &    0.16 &  40.2 &   2.8 &  \nodata &  \nodata \\
Cr I  & 5345.80 & 1.00 & $-$0.95 &  \nodata &  \nodata &  53.0 &   3.3 \\
Mn I  & 4030.75 & 0.00 & $-$0.48 &  72.6 &   2.8 &  \nodata &  \nodata \\
Mn I  & 4033.06 & 0.00 & $-$0.62 &  49.8 &   2.4 &  \nodata &  \nodata \\
Fe I  & 3565.38 & 0.96 & $-$0.13 & 104.4 &   4.7 &  \nodata &  \nodata \\
Fe I  & 3608.86 & 1.01 & $-$0.09 & 104.3 &   4.7 &  \nodata &  \nodata \\
Fe I  & 3727.62 & 0.96 & $-$0.61 &  90.3 &   4.7 &  \nodata &  \nodata \\
Fe I  & 3743.36 & 0.99 & $-$0.79 &  85.6 &   4.7 &  \nodata &  \nodata \\
Fe I  & 3753.61 & 2.18 & $-$0.89 &  32.8 &   4.8 &  \nodata &  \nodata \\
Fe I  & 3767.19 & 1.01 & $-$0.39 & 103.1 &   4.8 &  \nodata &  \nodata \\
Fe I  & 3786.68 & 1.01 & $-$2.19 &  20.9 &   4.6 &  \nodata &  \nodata \\
Fe I  & 3805.34 & 3.30 &    0.31 &  40.9 &   4.8 &  \nodata &  \nodata \\
Fe I  & 3827.82 & 1.56 &    0.09 &  99.3 &   4.7 &  \nodata &  \nodata \\
Fe I  & 3856.37 & 0.05 & $-$1.28 & 103.2 &   4.7 &  \nodata &  \nodata \\
Fe I  & 3902.95 & 1.56 & $-$0.44 &  78.9 &   4.7 &  \nodata &  \nodata \\
Fe I  & 3917.18 & 0.99 & $-$2.15 &  31.5 &   4.8 &  \nodata &  \nodata \\
Fe I  & 3940.88 & 0.96 & $-$2.60 &  19.1 &   4.9 &  \nodata &  \nodata \\
Fe I  & 4005.24 & 1.56 & $-$0.58 &  75.2 &   4.7 &  \nodata &  \nodata \\
Fe I  & 4014.53 & 3.05 & $-$0.59 &  15.7 &   4.8 &  \nodata &  \nodata \\
Fe I  & 4045.81 & 1.49 &    0.28 & 116.0 &   4.7 &  \nodata &  \nodata \\
Fe I  & 4063.59 & 1.56 &    0.06 & 103.7 &   4.7 &  \nodata &  \nodata \\
Fe I  & 4071.74 & 1.61 & $-$0.01 &  94.0 &   4.7 &  \nodata &  \nodata \\
Fe I  & 4132.06 & 1.61 & $-$0.68 &  64.2 &   4.6 &  \nodata &  \nodata \\
Fe I  & 4143.41 & 3.05 & $-$0.20 &  33.0 &   4.9 &  \nodata &  \nodata \\
Fe I  & 4143.87 & 1.56 & $-$0.51 &  72.4 &   4.6 &  \nodata &  \nodata \\
Fe I  & 4191.43 & 2.47 & $-$0.67 &  33.5 &   4.8 &  \nodata &  \nodata \\
Fe I  & 4202.03 & 1.49 & $-$0.69 &  69.4 &   4.6 &  \nodata &  \nodata \\
Fe I  & 4227.43 & 3.33 &    0.27 &  34.4 &   4.7 &  \nodata &  \nodata \\
Fe I  & 4250.79 & 1.56 & $-$0.71 &  69.3 &   4.7 &  \nodata &  \nodata \\
Fe I  & 4260.47 & 2.40 &    0.08 &  71.2 &   4.7 &  \nodata &  \nodata \\
Fe I  & 4404.75 & 1.56 & $-$0.15 &  96.5 &   4.7 &  \nodata &  \nodata \\
Fe I  & 4461.65 & 0.09 & $-$3.19 &  28.1 &   4.8 &  \nodata &  \nodata \\
Fe I  & 4466.55 & 2.83 & $-$0.60 &  28.2 &   4.9 &  \nodata &  \nodata \\
Fe I  & 4494.56 & 2.20 & $-$1.14 &  24.4 &   4.8 &  \nodata &  \nodata \\
Fe I  & 4528.61 & 2.18 & $-$0.82 &  34.9 &   4.7 &  \nodata &  \nodata \\
Fe I  & 4871.32 & 2.87 & $-$0.36 &  26.8 &   4.7 &  \nodata &  \nodata \\
Fe I  & 4872.14 & 2.88 & $-$0.57 &  20.3 &   4.7 &  \nodata &  \nodata \\
Fe I  & 4890.76 & 2.88 & $-$0.39 &  32.1 &   4.8 &  \nodata &  \nodata \\
Fe I  & 4891.49 & 2.85 & $-$0.11 &  47.8 &   4.8 &  \nodata &  \nodata \\
Fe I  & 4918.99 & 2.85 & $-$0.34 &  24.4 &   4.6 &  \nodata &  \nodata \\
Fe I  & 4920.50 & 2.83 &    0.07 &  42.7 &   4.5 &  \nodata &  \nodata \\
Fe I  & 5166.28 & 0.00 & $-$4.12 &  \nodata &  \nodata &  61.3 &   4.9 \\
Fe I  & 5171.60 & 1.49 & $-$1.72 &  34.4 &   4.8 &  92.6 &   4.8 \\
Fe I  & 5192.34 & 3.00 & $-$0.42 &  19.5 &   4.7 &  \nodata &  \nodata \\
Fe I  & 5194.94 & 1.56 & $-$2.02 &  \nodata &  \nodata &  75.4 &   4.9 \\
Fe I  & 5198.71 & 2.22 & $-$2.09 &  \nodata &  \nodata &  23.0 &   4.8 \\
Fe I  & 5202.34 & 2.18 & $-$1.87 &  \nodata &  \nodata &  52.4 &   5.0 \\
Fe I  & 5216.27 & 1.61 & $-$2.08 &  \nodata &  \nodata &  66.6 &   4.8 \\
Fe I  & 5217.39 & 3.21 & $-$1.16 &  \nodata &  \nodata &  31.0 &   5.1 \\
Fe I  & 5232.94 & 2.94 & $-$0.06 &  44.3 &   4.8 &  72.9 &   4.5 \\
Fe I  & 5254.96 & 0.11 & $-$4.76 &  \nodata &  \nodata &  21.4 &   4.9 \\
Fe I  & 5266.56 & 3.00 & $-$0.39 &  21.2 &   4.7 &  66.0 &   4.7 \\
Fe I  & 5269.54 & 0.86 & $-$1.33 &  81.7 &   4.8 & 153.5 &   4.9 \\
Fe I  & 5281.79 & 3.04 & $-$0.83 &  \nodata &  \nodata &  38.6 &   4.8 \\
Fe I  & 5283.62 & 3.24 & $-$0.52 &  \nodata &  \nodata &  50.3 &   4.9 \\
Fe I  & 5324.18 & 3.21 & $-$0.10 &  30.4 &   4.8 &  72.2 &   4.8 \\
Fe I  & 5328.04 & 0.92 & $-$1.47 &  64.6 &   4.6 & 128.8 &   4.7 \\
Fe I  & 5328.53 & 1.56 & $-$1.85 &  16.2 &   4.6 &  84.7 &   4.8 \\
Fe I  & 5371.49 & 0.96 & $-$1.64 &  61.8 &   4.8 & 142.9 &   5.2 \\
Fe I  & 5397.13 & 0.92 & $-$1.98 &  45.3 &   4.7 & 116.0 &   4.9 \\
Fe I  & 5405.77 & 0.99 & $-$1.85 &  47.4 &   4.7 & 100.4 &   4.5 \\
Fe I  & 5415.20 & 4.39 &    0.64 &  \nodata &  \nodata &  43.0 &   4.9 \\
Fe I  & 5429.70 & 0.96 & $-$1.88 &  48.2 &   4.7 & 106.9 &   4.6 \\
Fe I  & 5434.52 & 1.01 & $-$2.13 &  28.1 &   4.6 &  \nodata &  \nodata \\
Fe I  & 5446.92 & 0.99 & $-$1.91 &  45.3 &   4.7 &  \nodata &  \nodata \\
Fe I  & 5586.76 & 3.37 & $-$0.14 &  21.7 &   4.8 &  57.9 &   4.8 \\
Fe I  & 5615.64 & 3.33 &    0.05 &  21.2 &   4.5 &  \nodata &  \nodata \\
Fe I  & 5658.82 & 3.40 & $-$0.79 &  \nodata &  \nodata &  19.7 &   4.7 \\
Fe I  & 6065.48 & 2.61 & $-$1.41 &  \nodata &  \nodata &  30.0 &   4.6 \\
Fe I  & 6136.61 & 2.45 & $-$1.41 &  \nodata &  \nodata &  51.3 &   4.8 \\
Fe I  & 6137.69 & 2.59 & $-$1.35 &  \nodata &  \nodata &  44.7 &   4.8 \\
Fe I  & 6219.28 & 2.20 & $-$2.45 &  \nodata &  \nodata &  24.8 &   5.1 \\
Fe I  & 6230.72 & 2.56 & $-$1.28 &  \nodata &  \nodata &  43.8 &   4.7 \\
Fe I  & 6252.56 & 2.40 & $-$1.69 &  \nodata &  \nodata &  29.8 &   4.7 \\
Fe I  & 6393.60 & 2.43 & $-$1.58 &  \nodata &  \nodata &  57.2 &   5.0 \\
Fe I  & 6400.00 & 3.60 & $-$0.29 &  \nodata &  \nodata &  49.0 &   5.0 \\
Fe I  & 6421.35 & 2.28 & $-$2.01 &  \nodata &  \nodata &  28.8 &   4.8 \\
Fe I  & 6430.85 & 2.18 & $-$1.95 &  \nodata &  \nodata &  43.4 &   4.9 \\
Fe I  & 6494.98 & 2.40 & $-$1.24 &  \nodata &  \nodata &  60.3 &   4.7 \\
Fe I  & 6677.99 & 2.69 & $-$1.42 &  \nodata &  \nodata &  52.0 &   5.1 \\
Fe II & 4233.17 & 2.58 & $-$1.97 &  35.8 &   4.8 &  \nodata &  \nodata \\
Fe II & 4522.63 & 2.84 & $-$2.25 &  18.8 &   5.0 &  \nodata &  \nodata \\
Fe II & 4583.84 & 2.81 & $-$1.93 &  18.2 &   4.6 &  \nodata &  \nodata \\
Fe II & 4923.93 & 2.89 & $-$1.32 &  38.7 &   4.5 &  \nodata &  \nodata \\
Fe II & 5018.45 & 2.89 & $-$1.22 &  50.6 &   4.7 &  \nodata &  \nodata \\
Fe II & 5234.63 & 3.22 & $-$2.18 &  \nodata &  \nodata &  37.2 &   4.8 \\
Fe II & 5276.00 & 3.20 & $-$2.01 &  \nodata &  \nodata &  49.3 &   4.9 \\
Co I  & 3845.47 & 0.92 &    0.01 &  60.0 &   3.0 &  \nodata &  \nodata \\
Co I  & 4121.32 & 0.92 & $-$0.32 &  24.5 &   2.6 &  \nodata &  \nodata \\
Ni I  & 3807.14 & 0.42 & $-$1.22 &  47.7 &   3.3 &  \nodata &  \nodata \\
Ni I  & 3858.30 & 0.42 & $-$0.95 &  65.0 &   3.5 &  \nodata &  \nodata \\
Zn I  & 4722.15 & 4.03 & $-$0.39 &  6.5\tablenotemark{a} &  2.5 & \nodata &  \nodata \\
Zn I  & 4810.53 & 4.08 & $-$0.14 &  4.4\tablenotemark{a} &  2.1 &  13.3\tablenotemark{a} &   1.9 \\ 
Sr   II   & 4077.00 &   0.00 &  $-$1.26 & syn &  0.4 & \nodata & \nodata \\ 
Sr   II   & 4215.00 &   0.00 &  $-$1.32 & syn &  0.3 & \nodata & \nodata \\ 
Sr   II   & 4607.33 &   0.00 &  $-$0.57 & \nodata & \nodata & syn &  1.0 \\ 
Y    II   & 3774.33 &   0.13 &   0.21 & syn & $-$0.1 & \nodata & \nodata \\ 
Y    II   & 4854.87 &   0.99 &  $-$0.38 & \nodata & \nodata & syn &  0.2 \\ 
Y    II   & 5200.41 &   0.99 &  $-$0.57 & \nodata & \nodata & syn &  0.2 \\ 
Y    II   & 5205.73 &   1.03 &  $-$0.34 & \nodata & \nodata & syn &  0.3 \\ 
Zr   II   & 4050.33 &   0.71 &  $-$1.00 & \nodata & \nodata & syn &  0.6 \\ 
Zr   II   & 4208.99 &   0.71 &  $-$0.46 & syn &  0.6 & \nodata & \nodata \\ 
Ba   II   & 4554.03 &   0.00 &   0.14 & syn &  1.3 & \nodata & \nodata \\ 
Ba   II   & 4934.10 &   0.00 &  $-$0.16 & syn &  1.4 & \nodata & \nodata \\ 
Ba   II   & 5853.68 &   0.60 &  $-$2.56 & syn &  1.3 & syn &  1.6 \\ 
Ba   II   & 6141.71 &   0.70 &  $-$0.08 & syn &  1.2 & syn &  1.7 \\ 
Ba   II   & 6496.91 &   0.60 &  $-$0.38 & syn &  1.3 & syn &  1.9 \\ 
La   II   & 3995.74 &   0.17 &  $-$0.06 & syn &  0.0 & \nodata & \nodata \\ 
La   II   & 4086.71 &   0.00 &  $-$0.07 & syn &  0.0 & \nodata & \nodata \\ 
La   II   & 4123.22 &   0.32 &   0.13 & syn &  0.0 & \nodata & \nodata \\ 
La   II   & 4526.12 &   0.77 &  $-$0.59 & \nodata & \nodata & syn &  0.3 \\ 
La   II   & 4921.79 &   0.24 &  $-$0.45 & \nodata & \nodata & syn &  0.3 \\ 
La   II   & 5290.84 &   0.00 &  $-$1.65 & \nodata & \nodata & syn &  0.3 \\ 
La   II   & 5303.53 &   0.32 &  $-$1.35 & \nodata & \nodata & syn &  0.4 \\ 
La   II   & 5797.57 &   0.24 &  $-$1.36 & \nodata & \nodata & syn &  0.4 \\ 
Ce   II   & 4053.50 &   0.00 &  $-$0.71 & syn &  0.7 & \nodata & \nodata \\ 
Ce   II   & 4083.22 &   0.70 &   0.27 & syn &  0.5 & \nodata & \nodata \\ 
Ce   II   & 4127.36 &   0.68 &   0.31 & \nodata & \nodata & syn &  0.5 \\ 
Ce   II   & 4222.64 &   0.79 &  $-$1.31 & \nodata & \nodata & syn &  0.5 \\ 
Ce   II   & 4562.36 &   0.48 &   0.23 & syn &  0.6 & syn &  0.7 \\ 
Pr   II   & 3964.82 &   0.06 &   0.12 & \nodata & \nodata & syn & $-$0.3 \\ 
Pr   II   & 3965.26 &   0.20 &   0.14 & \nodata & \nodata & syn & $-$0.1 \\ 
Pr   II   & 5220.01 &   0.79 &   0.30 & \nodata & \nodata & syn &  0.0 \\ 
Pr   II   & 5292.61 &   0.65 &  $-$0.26 & \nodata & \nodata & syn &  0.0 \\ 
Nd   II   & 4004.00 &   0.06 &  $-$0.57 & \nodata & \nodata & syn &  0.5 \\ 
Nd   II   & 4011.06 &   0.47 &  $-$0.76 & \nodata & \nodata & syn &  0.5 \\ 
Nd   II   & 4012.70 &   0.00 &  $-$0.60 & \nodata & \nodata & syn &  0.5 \\ 
Nd   II   & 4013.22 &   0.18 &  $-$1.10 & \nodata & \nodata & syn &  0.5 \\ 
Nd   II   & 4021.33 &   0.32 &  $-$0.10 & syn &  0.2 & syn &  0.3 \\ 
Nd   II   & 4043.59 &   0.32 &  $-$0.71 & \nodata & \nodata & syn &  0.5 \\ 
Nd   II   & 4061.08 &   0.47 &   0.55 & syn &  0.2 & syn &  0.6 \\ 
Nd   II   & 4069.26 &   0.06 &  $-$0.57 & \nodata & \nodata & syn &  0.6 \\ 
Nd   II   & 5249.58 &   0.97 &   0.20 & \nodata & \nodata & syn &  0.6 \\ 
Sm   II   & 4318.93 &   0.28 &  $-$0.25 & syn & $-$0.3 & \nodata & \nodata \\ 
Sm   II   & 4434.32 &   0.38 &  $-$0.07 & \nodata & \nodata & syn &  0.0 \\ 
Sm   II   & 4467.34 &   0.66 &   0.15 & syn & $-$0.3 & \nodata & \nodata \\ 
Sm   II   & 4499.48 &   0.25 &  $-$0.87 & \nodata & \nodata & syn &  0.2 \\ 
Eu   II   & 3724.93 &   0.00 &  $-$0.09 & \nodata & \nodata & syn & $-$0.4 \\ 
Eu   II   & 3907.11 &   0.21 &   0.17 & \nodata & \nodata & syn & $-$0.6 \\ 
Eu   II   & 4129.72 & 0.00 &    0.22 &   8.6\tablenotemark{a} &$-$1.1 & \nodata &  \nodata \\
Eu   II   & 4205.04 & 0.00 &    0.21 &   6.5\tablenotemark{a} &$-$1.3 & \nodata &  \nodata \\
Eu   II   & 6645.06 &   1.38 &   0.12 & \nodata & \nodata & syn & $-$0.4 \\ 
Gd   II   & 3481.80 &   0.49 &   0.11 & \nodata & \nodata & syn &  0.1 \\ 
Gd   II   & 4251.73 &   0.38 &  $-$0.22 & \nodata & \nodata & syn &  0.2 \\ 
Tb   II   & 3702.85 &   0.13 &   0.44 & \nodata & \nodata & syn & $-$0.7 \\ 
Dy   II   & 3445.57 &   0.00 &  $-$0.15 & \nodata & \nodata & syn &  0.2 \\ 
Dy   II   & 3531.71 &   0.00 &   0.77 & \nodata & \nodata & syn &  0.2 \\ 
Dy   II   & 4103.31 &   0.10 &  $-$0.38 & \nodata & \nodata & syn &  0.0 \\ 
Er   II   & 3729.52 &   0.00 &  $-$0.59 & \nodata & \nodata & syn &  0.0 \\ 
Er   II   & 3938.63 &   0.00 &  $-$0.52 & \nodata & \nodata & syn &  0.0 \\ 
Tm   II   & 3700.26 &   0.03 &  $-$0.38 & \nodata & \nodata & syn & $-$0.9 \\ 
Tm   II   & 3701.36 &   0.00 &  $-$0.54 & \nodata & \nodata & syn & $-$0.8 \\ 
Yb   II   & 3694.37 &   0.00 &  $-$0.30 & syn & $-$1.0 & \nodata & \nodata \\ 
Hf   II   & 3918.08 &   0.45 &  $-$1.01 & syn &  0.2 & syn &  0.1 \\ 
Hf   II   & 4093.15 &   0.45 &  $-$1.15 & \nodata & \nodata & syn &  0.3 \\ 
Os   I    & 4260.80 &   0.00 &  $-$1.44 & \nodata & \nodata & syn &  0.1 \\ 
Pb   I    & 3683.46 &   0.97 &  $-$0.46 & syn &  2.5 & syn &  1.9 \\ 
Pb   I    & 4057.81 &   1.32 &  $-$0.17 & syn &  2.5 & syn &  1.9 \\
\enddata 
\tablenotetext{a}{Upper limits.}
\end{deluxetable}

\begin{deluxetable}{lcrrrrrrcrrrrr} 
\tabletypesize{\scriptsize}
\tablecaption{\label{abund1} Abundances}   
\tablewidth{0pt}  
\tablehead{  
\colhead{}&\colhead{}&\colhead{}& 
\multicolumn{5}{c}{HE~2138$-$3336}& \colhead{}& \multicolumn{5}{c}{HE~2258$-$6358}\\ 
\cline{4-8} 
\cline{10-14}\\ 
\colhead{Element} &  
\colhead{Ion} &  
\colhead{$\log\epsilon$\,(X)$_{\odot}$}&  
\colhead{$\log\epsilon$\,(X)} & 
\colhead{$\mbox{[X/H]}$}&  
\colhead{$\mbox{[X/Fe]}$}&  
\colhead{$\sigma$} &
\colhead{$N_{\rm lines}$}& 
\colhead{}& 
\colhead{$\log\epsilon$\,(X)} & 
\colhead{$\mbox{[X/H]}$}&  
\colhead{$\mbox{[X/Fe]}$}&  
\colhead{$\sigma$} &
\colhead{$N_{\rm lines}$} }
\startdata  
C  & \nodata & 8.43 & 8.08 & $-$0.36 &    2.43  &    0.05  &  4        &&    8.18  & $-$0.25  &    2.42 &    0.09 &       4 \\
N  & \nodata & 7.83 & 6.70 & $-$1.13 &    1.66  &    0.20  &  1        &&    6.60  & $-$1.23  &    1.44 &    0.25 &       1 \\
O  & 1  & 8.69 & \nodata  & \nodata  & \nodata  & \nodata  & \nodata   &&    7.87  & $-$0.82  &    1.85 &    0.10 &       1 \\
Na & 1  & 6.24 & 4.67     & $-$1.57  &    1.22  &    0.10  &  2        &&    4.22  & $-$2.02  &    0.65 &    0.10 &       2 \\
Mg & 1  & 7.60 & 5.38     & $-$2.22  &    0.57  &    0.12  &  3        &&    5.65  & $-$1.95  &    0.72 &    0.30 &       2 \\
Al & 1  & 6.45 & 3.21     & $-$3.24  & $-$0.45  &    0.10  &  1        &&    2.90  & $-$3.55  & $-$0.88 &    0.10 &       1 \\
Si & 1  & 7.51 & \nodata  & \nodata  & \nodata  & \nodata  & \nodata   &&    5.20  & $-$2.31  &    0.38 &    0.10 &       1 \\
Ca & 1  & 6.34 & 3.94     & $-$2.40  &    0.39  &    0.09  &  4        &&    4.32  & $-$2.02  &    0.65 &    0.09 &       3 \\
Sc & 2  & 3.15 & 1.23     & $-$1.92  &    0.87  &    0.10  &  1        &&    0.79  & $-$2.36  &    0.31 &    0.05 &       2 \\
Ti & 1  & 4.95 & \nodata  & \nodata  & \nodata  & \nodata  & \nodata   &&    2.79  & $-$2.16  &    0.51 &    0.10 &       1 \\
Ti & 2  & 4.95 & 2.40     & $-$2.55  &    0.24  &    0.04  &  8        &&    2.84  & $-$2.11  &    0.56 &    0.08 &       3 \\
Cr & 1  & 5.64 & 2.76     & $-$2.88  & $-$0.09  &    0.07  &  3        &&    3.14  & $-$2.50  &    0.17 &    0.16 &       2 \\
Mn & 1  & 5.43 & 2.56     & $-$2.87  & $-$0.08  &    0.24  &  2        && \nodata  & \nodata  & \nodata & \nodata & \nodata \\
Fe & 1  & 7.50 & 4.71     & $-$2.79  &    0.00  &    0.01  & 53        &&    4.83  & $-$2.67  &    0.00 &    0.03 &      35 \\
Fe & 2  & 7.50 & 4.71     & $-$2.79  &    0.00  &    0.13  &  5        &&    4.85  & $-$2.65  &    0.02 &    0.03 &       2 \\
Co & 1  & 4.99 & 2.81     & $-$2.18  &    0.61  &    0.29  &  2        && \nodata  & \nodata  & \nodata & \nodata & \nodata \\
Ni & 1  & 6.22 & 3.40     & $-$2.82  & $-$0.03  &    0.08  &  2        && \nodata  & \nodata  & \nodata & \nodata & \nodata \\
Zn & 1  & 4.56 & 2.30     & $-$2.26  & $<$0.53  & \nodata  &  2        &&    1.94  & $-$2.62  & $<$0.05 & \nodata &       1 \\

Sr & 2  & 2.87 &    0.35  & $-$2.52  &    0.27  &    0.15  &  2        &&      1.00 & $-$1.87 &    0.80 &    0.15 &       1 \\
Y  & 2  & 2.21 & $-$0.10  & $-$2.31  &    0.48  &    0.15  &  1        &&      0.24 & $-$1.97 &    0.70 &    0.10 &       3 \\
Zr & 2  & 2.58 &    0.60  & $-$1.98  &    0.81  &    0.20  &  1        &&      0.60 & $-$1.98 &    0.69 &    0.15 &       1 \\
Ba & 2  & 2.18 &    1.30  & $-$0.88  &    1.91  &    0.07  &  5        &&      1.74 & $-$0.44 &    2.23 &    0.11 &       3 \\
La & 2  & 1.10 &    0.00  & $-$1.10  &    1.60  &    0.10  &  3        &&      0.34 & $-$0.76 &    1.91 &    0.09 &       5 \\
Ce & 2  & 1.58 &    0.60  & $-$0.98  &    1.81  &    0.12  &  3        &&      0.57 & $-$1.01 &    1.66 &    0.10 &       3 \\
Pr & 2  & 0.72 & \nodata  & \nodata  & \nodata  & \nodata  & \nodata   &&   $-$0.10 & $-$0.82 &    1.85 &    0.10 &       4 \\
Nd & 2  & 1.42 &    0.20  & $-$1.22  &    1.57  &    0.15  &  2        &&      0.51 & $-$0.91 &    1.76 &    0.04 &       9 \\
Sm & 2  & 0.96 & $-$0.30  & $-$1.26  &    1.53  &    0.10  &  2        &&      0.10 & $-$0.86 &    1.81 &    0.13 &       2 \\
Eu & 2  & 0.52 & $-$1.18  & $-$1.70  & $<$1.09  & \nodata  &  2        &&   $-$0.47 & $-$0.99 &    1.68 &    0.07 &       3 \\
Gd & 2  & 1.07 & \nodata  & \nodata  & \nodata  & \nodata  & \nodata   &&      0.15 & $-$0.92 &    1.75 &    0.09 &       2 \\
Tb & 2  & 0.30 & \nodata  & \nodata  & \nodata  & \nodata  & \nodata   &&   $-$0.70 & $-$1.00 &    1.67 &    0.15 &       1 \\
Dy & 2  & 1.10 & \nodata  & \nodata  & \nodata  & \nodata  & \nodata   &&      0.03 & $-$1.07 &    1.60 &    0.10 &       4 \\
Er & 2  & 0.92 & \nodata  & \nodata  & \nodata  & \nodata  & \nodata   &&      0.00 & $-$0.92 &    1.75 &    0.15 &       2 \\
Tm & 2  & 0.10 & \nodata  & \nodata  & \nodata  & \nodata  & \nodata   &&   $-$0.88 & $-$0.98 &    1.69 &    0.15 &       2 \\
Yb & 2  & 0.84 & $-$1.00  & $-$1.84  &    0.95  &    0.25  &  1        &&   \nodata & \nodata & \nodata & \nodata & \nodata \\
Hf & 2  & 0.85 &    0.20  & $-$0.65  &    2.14  &    0.20  &  1        &&      0.20 & $-$0.65 &    2.02 &    0.15 &       2 \\
Os & 1  & 1.40 & \nodata  & \nodata  & \nodata  & \nodata  & \nodata   &&      0.10 & $-$1.30 &    1.37 &    0.20 &       1 \\
Pb & 1  & 1.75 &    2.50  &    0.75  &    3.54\tablenotemark{a}  &    0.15  &  2        &&      1.90 &    0.15 &    2.82\tablenotemark{b} &    0.15 &       2   

\enddata      
\tablenotetext{a}{3.84 including NLTE correction}
\tablenotetext{b}{3.32 including NLTE correction}
\end{deluxetable}

\begin{deluxetable}{lrrr}
\tabletypesize{\scriptsize}
  \tablewidth{200pt}
  \tablecaption{Example Systematic Abundance Uncertainties for HE~2138$-$3336 \label{syserrstab}}
  \tablehead{\colhead{Elem}&\colhead{$\Delta$\teff}&\colhead{$\Delta$\logg}&\colhead{$\Delta\xi$}\\
\colhead{}&\colhead{$+$150\,K}&\colhead{$+$0.5 dex}&\colhead{$+$0.3 km/s}}
\startdata
Na I  & 0.19 &    0.24 & 0.07 \\
Mg I  & 0.11 &    0.10 & 0.03 \\
Al I  & 0.15 &    0.09 & 0.09 \\
Ca I  & 0.09 &    0.02 & 0.02 \\
Sc II & 0.11 & $-$0.13 & 0.18 \\
Ti II & 0.07 & $-$0.16 & 0.07 \\
Cr I  & 0.15 &    0.02 & 0.05 \\
Mn I  & 0.18 &    0.03 & 0.10 \\
Fe I  & 0.15 &    0.04 & 0.07 \\
Fe II & 0.02 & $-$0.18 & 0.03 \\
Co I  & 0.15 &    0.00 & 0.05 \\
Ni I  & 0.17 &    0.01 & 0.09 \\
Sr II & 0.14 & $-$0.04 & 0.22 \\
Ba II & 0.16 &    0.03 & 0.14

\enddata
\end{deluxetable}

\begin{deluxetable}{lccccccccccr}
\tabletypesize{\scriptsize}
  \tablewidth{0pc}
  \tablecaption{Data for Literature Stars\label{other_stars}}
  \tablehead{
\colhead{Star}&
\colhead{Type}&
\colhead{\metal}&
\colhead{\teff}&
\colhead{\logg}&
\colhead{N\tablenotemark{a}}&
\multicolumn{2}{c}{$M_\mathrm{i}~(M_{\odot})$=0.00\tablenotemark{b}}&
\colhead{}&
\multicolumn{2}{c}{$M_\mathrm{i}~(M_{\odot})$=0.38\tablenotemark{b}}&
\colhead{Ref.\tablenotemark{c}}\\
\cline{7-8} 
\cline{10-11}\\
\colhead{}&
\colhead{}&
\colhead{(dex)}&
\colhead{(K)}&
\colhead{(cgs)}&
\colhead{}&
\colhead{TP\#}&
\colhead{RES}&
\colhead{}&
\colhead{TP\#}&
\colhead{RES}&
\colhead{}
}
\startdata

   CS~22183-015  &  s    &  -3.00  &  5200  &  2.50  &  11  &   5  &  0.39  &&   6  &  0.09  &  1  \\
   CS~22898-027  &  r/s  &  -2.25  &  6250  &  3.70  &  12  &  15  &  0.25  &&  21  &  0.57  &  2  \\
   CS~22942-019  &  s    &  -2.64  &  5000  &  2.40  &  11  &   8  &  0.54  &&  15  &  0.31  &  2  \\
   CS~22947-187  &  s    &  -2.47  &  5160  &  1.30  &  10  &   5  &  0.57  &&   5  &  0.16  &  3  \\
   CS~22948-027  &  r/s  &  -2.47  &  4800  &  1.80  &  11  &  10  &  0.15  &&  21  &  0.21  &  4  \\
   CS~22964-161  &  s    &  -2.39  &  6050  &  3.70  &  11  &   5  &  0.32  &&   8  &  0.06  &  5  \\
   CS~29497-030  &  r/s  &  -2.57  &  7000  &  4.10  &  19  &  15  &  0.37  &&  21  &  0.70  &  6  \\
   CS~29497-034  &  r/s  &  -2.90  &  4800  &  1.80  &  11  &   5  &  0.09  &&  13  &  0.07  &  5  \\
   CS~29526-110  &  r/s  &  -2.38  &  6500  &  3.20  &   9  &  10  &  0.11  &&  21  &  0.22  &  2  \\
   CS~30301-015  &  s    &  -2.64  &  4750  &  0.80  &  11  &   5  &  0.75  &&   5  &  0.12  &  2  \\
   CS~31062-012  &  r/s  &  -2.55  &  6250  &  4.50  &   9  &   5  &  0.16  &&  13  &  0.26  &  2  \\
   CS~31062-050  &  r/s  &  -2.31  &  5500  &  2.70  &  21  &  14  &  0.20  &&  21  &  0.47  &  7  \\
      HD~196944  &  s    &  -2.25  &  5250  &  1.80  &  13  &   5  &  0.27  &&   9  &  0.08  &  2  \\
   HE~0058-0244  &  r/s  &  -2.75  &  5730  &  3.50  &   9  &   5  &  0.18  &&   9  &  0.20  &  8  \\
   HE~0131-3953  &  r/s  &  -2.50  &  6322  &  3.85  &   7  &   5  &  0.07  &&  14  &  0.16  &  9  \\
   HE~0143-0441  &  s    &  -2.31  &  6240  &  3.70  &  10  &   9  &  0.12  &&  21  &  0.21  &  8  \\
   HE~0202-2204  &  s    &  -1.98  &  5621  &  3.47  &  10  &   5  &  0.09  &&  14  &  0.12  &  9  \\
   HE~0338-3945  &  r/s  &  -2.43  &  6160  &  4.13  &  21  &  21  &  0.46  &&  21  &  0.90  &  10  \\
   HE~1031-0020  &  s    &  -2.86  &  5080  &  2.20  &   9  &   5  &  0.61  &&   5  &  0.07  &  8  \\
   HE~1105+0027  &  r/s  &  -2.42  &  6121  &  3.75  &   7  &   8  &  0.25  &&  21  &  0.39  &  9  \\
   HE~1135+0139  &  s    &  -2.31  &  5736  &  3.55  &   9  &   5  &  0.28  &&   8  &  0.07  &  9  \\
   HE~1509-0806  &  s    &  -2.91  &  5185  &  2.50  &   9  &   5  &  0.22  &&   9  &  0.08  &  8  \\
   HE~2148-1247  &  r/s  &  -2.37  &  6380  &  3.90  &  14  &  14  &  0.30  &&  21  &  0.60  &  11  \\
   HE~2158-0348  &  s    &  -2.70  &  5215  &  2.50  &  11  &   5  &  0.24  &&  11  &  0.21  &  8  \\
 HKII~17435-00532  &  s    &  -2.23  &  5200  &  2.15  &  11  &   5  &  0.28  &&   8  &  0.12  &  12  \\
      LP~625-44  &  r/s  &  -2.71  &  5500  &  2.80  &  17  &  11  &  0.23  &&  21  &  0.34  &  2 
 
\enddata
\tablenotetext{a}{Number of observed abundances.}
\tablenotetext{b}{$M_\mathrm{acc}~(M_{\odot})$=0.10.}
\tablenotetext{c}{1. \citet{johnson2002}, 2. \citet{aoki2002}, 3. \citet{mcw1995}, 4. \citet{barbuy2005}, 5. \citet{thompson2008}, 6. \citet{ivans2005}, 7. \citet{johnson2004}, 8. \citet{cohen2006}, 9. \citet{barklem2005}, 10. \citet{jonsell2006}, 11. \citet{cohen2003}, 12. \citet{roederer2008}.}
\end{deluxetable}

\end{document}